\documentclass[sn-mathphys,Numbered]{sn-jnl}

\usepackage{graphicx}%
\usepackage[caption=false,font=normalsize,labelfont=sf,textfont=sf]{subfig}
\usepackage{multirow}%
\usepackage{amsmath,amssymb,amsfonts}%
\usepackage{amsthm}%
\usepackage{mathrsfs}%
\usepackage[title]{appendix}%
\usepackage{xcolor}%
\usepackage{textcomp}%
\usepackage{manyfoot}%
\usepackage{booktabs}%
\usepackage{algorithm}%
\usepackage{algorithmicx}%
\usepackage{algpseudocode}%
\usepackage{listings}%

\begin{document}

\title[Citizen Science Games on the Timeline of Quantum Games]{Citizen Science Games on the Timeline of Quantum Games}


\author[1,2]{\fnm{Laura} \sur{Piispanen}}\email{laura.piispanen@aalto.fi}

\affil[1]{\orgdiv{Department of Computer Science}, \orgname{Aalto University School of Science}, \orgaddress{\street{Konemiehentie 2}, \city{Espoo}, \postcode{FI-00076}, \country{Finland}}}
\affil[2]{\orgdiv{Department of Applied Physics}, \orgname{Aalto University School of Science}, \orgaddress{\street{Tietotie 3}, \city{Espoo}, \postcode{FI-00076}, \country{Finland}}}


\abstract{This article provides an overview of existing quantum physics-related games, referred to as \textit{quantum games}, that serve citizen science research in quantum physics. Additionally, we explore the connection between citizen science and \textit{quantum computer games}, games played on quantum computers. The information presented is derived from academic references and supplemented by diverse sources, including social media publications, conference presentations, and blog posts from research groups and developers associated with the presented games. We observe that the current landscape of quantum games is shaped by three distinct driving forces: the serious application of games, the evolution of quantum computers, and open game development events such as \textit{Quantum Game Jams}. Notably, citizen science plays an influential role in all three aspects. The article points to existing design guides for citizen science quantum games and views future prospects of citizen science projects and quantum games through collaborative endeavours, human-machine collaboration, and open access quantum computers.}




\maketitle

\section{Introduction}\label{sec1}
The popularity of digital games as a pastime activity for over 3 billion people \cite{gamers} 
has served as inspiration for applications in citizen science, education, and training. \textit{Citizen science} is the concept of involving members of the general public in scientific studies or research \cite{haklay2015, bonney2014, heigl2019, vohland2021}, with \textit{citizen science games} offering an interactive tool for such purposes \cite{cooper2011phd,cooper2018, pelacho2021}. Digital games may be used for collecting and processing data or even for learning and applying a domain skill \cite{miller2022barr}. Today, games are played on a diverse collection of platforms, including computers, mobile phones, specialised gaming hardware, and physical formats like board games and card games. The same applies to games that relate to \textit{quantum physics} \cite{piispanen2022,piispanen2023history}.

Quantum physics describes the phenomena and structure of the sub-microscopic scales governed by particles such as atoms, electrons, protons, and photons. It delves into the inherently probabilistic phenomena beyond our daily observation \cite{busch1995, zeilinger1999}. Despite seeming like an abstract theory, quantum physics underpins the technology behind the \textit{first generation quantum technologies}, such as lasers, transistors, and superconductors, which enable the existence of compact devices like mobile phones and smartwatches. Nowadays \textit{second generation quantum technologies}, with a specific focus on quantum computers and the field of quantum computing, are at the forefront of development \cite{dowling2003, deutsch2020}. 

Today, there exists over three hundred games that reference quantum physics, quantum computing, or quantum technologies \cite{piispanen2022, piispanen2023history, quantumgames}. \textit{Quantum games} are defined as any type of rule-based games that use the principles of, or reference the theory of, quantum physics or quantum phenomena through any of three \textit{dimensions of quantum games}: the perceivable dimension of quantum physics, the dimension of quantum technologies, and the dimension of scientific purposes like citizen science \cite{piispanen2022}. This definition encompasses games inspired by quantum physics for entertainment purposes, games designed for educational use in the teaching of quantum physics or quantum computing, citizen science games, and those developed specifically for quantum computers (See Table \ref{table:definition}) \cite{piispanen2022}.
\begin{table}[!ht]
\caption{The dimensions of quantum games as described by Piispanen et al. (2022) \cite{piispanen2022}.}
    \centering
    \begin{tabular}{|l|l||}
    \hline
        \textbf{DIMENSIONS OF QUANTUM GAMES} \\ \hline \hline
        \textbf{Perceivable dimension of quantum physics:} \\ 
        The reference to quantum physics in the game is perceivable by interacting \\
        with the game or with its peripheral material \\
        (such as rule books, descriptions, etc.) \\ \hline
        \textbf{Dimension of quantum technologies:} \\
        The game incorporates usage of quantum software or quantum devices\\ 
        either during the gameplay itself, or during the development of the game \\ \hline
        \textbf{Dimension of scientific purposes:} \\
        The game is intended to be an educational game, a citizen science game, \\
        uses a tool designed for such games or otherwise has a purpose \\
        towards a scientific use \\ \hline
    \end{tabular}
    \end{table}
\label{table:definition}

Most of the previous papers on quantum games have concentrated on educational quantum games or for other reasons have omitted examining citizen science quantum games \cite{goff2006,cantwell2019,anupam2020,piispanen2022, carberry2022, seskir2022, piispanen2023history, kopf2023endless}. In this article we review the history of quantum games from the perspective of \textit{citizen science quantum games} and the different motivations behind their development. This means that we will focus on citizen science games and citizen science game prototypes related to the study of quantum sciences, quantum technology development and the benchmarking of early quantum computers. In addition to academic publications, the references include blog posts and social media material disseminated by research teams associated with the presented citizen science quantum games. 

\section{Citizen Science and Gamification}
Citizen science has a long history of involving members of the general public as collaborators or researchers in research projects, thereby influencing information development and facilitating large-scale data gathering \cite{raddick2009citizen,eitzel2017,vohland2021, frigerio2021}. The problems addressed usually involve tasks that cannot be fully automated and/or tasks in which human abilities such as pattern recognition, and spatial reasoning are seen as preferable to algorithms or machine learning methods \cite{cooper2011phd,vohland2021}. \textit{Crowdsourcing}, the outsourcing of tasks to a large group of people or a community, typically through an online platform, serves as a means to facilitate and enhance the participation of the public in these endeavours \cite{wiggins2011}. In addition, citizen science has offered insight and knowledge about problem solving itself \cite{keep2018}. However, consensus on the term has not always been uniform and depending on the protocols and training provided to citizen scientists, the practice has faced criticism regarding the quality of data, scientific rigour, open-source data management technologies, and the data analysis tools employed by non-experts \cite{bonney2014, eitzel2017, heigl2019}. To address the criticism, project evaluation tools, formal data policies, data management and quality control plans have been developed for use in citizen science projects \cite{bonney2014}. For research teams, it proves helpful to study and adapt based on similar projects, which may share similarities in the form of the collected data or the research problem, when they are reported using templates relying on these policies.\\

The availability and affordability of electrical devices for a growing number of the general public has further enabled the growth of citizen science research and allowed for the collection of a multitude of data types. Access to the internet, computers and smartphones, in particular, have also further enabled the use of \textit{gamified} elements in data manipulation platforms for participating members. \textit{Gamification} is the application of game-design elements and principles in non-game contexts to enhance user engagement, productivity, learning, or other behaviours \cite{deterding2011, hamari2019}. These elements include for example granting points and rewards, fostering competition, leaderboards, and achievements and can be incorporated into any citizen science platform or forum to foster motivation and provide a rewarding environment for the task. Gamification is used for the purpose of motivating the tasks and commitment of the public both in terms of training the user as well as data gathering and/or manipulation \cite{sullivan2009,bowser2013using,bonney2014}. A fully gamified citizen science interface, that incorporates these elements within a single platform or application is called a \textit{citizen science game}: a fully self-standing game enabling the public-sourced tasks related to the citizen science project \cite{miller2022}. 

\textit{Foldit}, a citizen science game for protein structure manipulation launched in 2008, marked a significant milestone in citizen science research by enlisting players to solve computationally challenging protein-folding problems \cite{foldit,cooper2011phd} and later proved to be a tool for professional scientists themselves \cite{cooper2018}. This success has since spurred a wave of gamified initiatives spanning diverse scientific disciplines, from astronomy to biology \cite{raddick2009citizen,phylo2012,eterna2014,eyewire2017}. The use of citizen science games has gained prominence due to games' ability to harness the collective power of large and diverse participant groups for crowdsourcing in science. Games have emerged as powerful tools in the realm of citizen science as they offer an engaging platform for individuals to contribute to scientific research while enjoying a recreational activity. Games can bridge the gap between scientific research and public participation, offering a unique avenue for collaborative discovery and learning \cite{ bonney2014, heigl2019, vohland2021}.

In physics research citizen science plays a transformative role in breaking down barriers and fostering inclusivity \cite{citizenscience, yadav2018,sanchez2011}. Zooniverse.org is a large online platform for citizen science projects, which have resulted in over a hundred research papers within the last few years, also in the research field of physics \cite{simpson2014,zooniversephysics}. Actors like the National Aeronautics and Space Administration (NASA) provide another platform for citizen science games \cite{nasacisci}, but physical sciences are also present through the global citizen science hub \textit{SciStarter} \cite{scistarter}. In the next section we will concentrate on citizen science games that are developed for the study of quantum physical sciences, \textit{citizen science quantum games}. 

\section{Quantum Games and Citizen Science} 
In this section we review the history of citizen science quantum games alongside the history of quantum games, which is presented in references \cite{wootton2018GameOn,piispanen2022,piispanen2023history}. We concentrate on three different aspects that have bridged quantum games with citizen science under respective subsections. Many of the oldest games recorded that were using the word `quantum' or other terms related to quantum physics in their names have not necessarily had any further connection to quantum physics \cite{piispanen2022, piispanen2023history}. Instead, these terms and references have been used to boost the sci-fi theme of the game and create fantasy, thereby raising the concern of creating misconceptions \cite{piispanen2022}. This may be partly due to quantum physics education being limited mostly to university level studies in the past, but also to the perceived challenging nature of the theory. 

The era of online education and particularly the attention surrounding second generation quantum technologies has resulted in the creation of many curated online materials suitable for quantum education. In particular, educational games on quantum physics have taken the opportunity to incorporate simulated quantum phenomena into game mechanics and offer an attractive alternative or companion to traditional methods of quantum education \cite{piispanen2022,seskir2022,piispanen2023history}. The use of games in quantum education is not yet extensively studied, but some interactive tools that visualise quantum physical phenomena have proven to be successful in strengthening the learning and mental model building process related to these otherwise abstract concepts \cite{kohnle2010, kohnle2012, passante2019}. The development of educational games and learning platforms integrating games with learning materials has inspired many recent projects \cite{scienceathomegames, qplaylearn, helloquantum, seskir2022}.

\textit{Serious games} refer to games that are developed for education, training, citizen science or for another purpose that is not just entertainment \cite{abt1970,sawyer2007,ratan2009,djaouti2011,nagarajan2012}. In the following we will concentrate on particularly on citizen science games. Though this section concentrates on digital games, it's worth noting that analogue quantum games, like card and board games, also exist \cite{piispanen2022, piispanen2023history, quantumgames}. No examples of analogue quantum games for the purposes of citizen science were found in this study.
\begin{figure}[ht]
\center
\subfloat[]{\includegraphics[width = 0.3\linewidth]{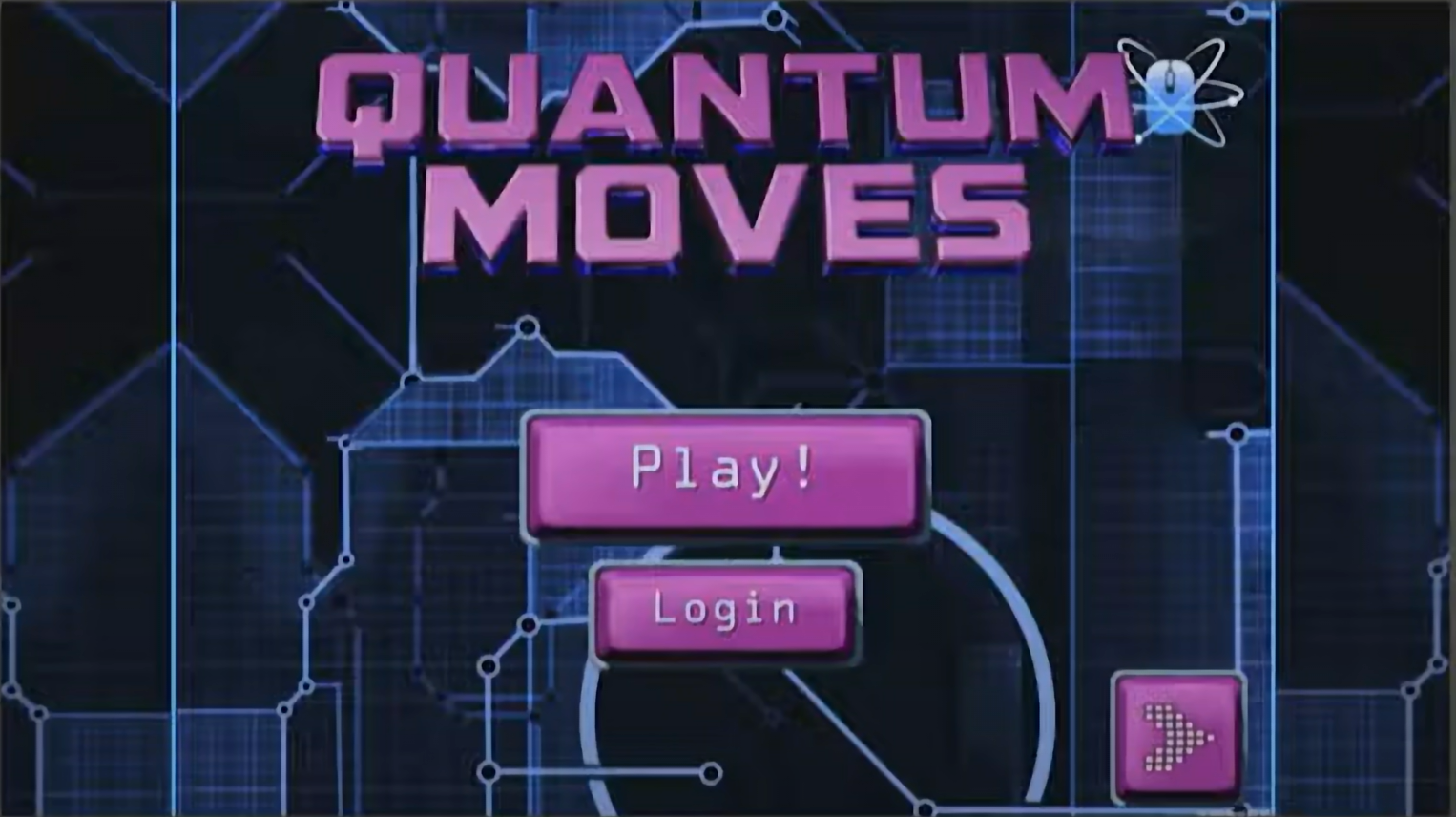}\,
\includegraphics[width = 0.3\linewidth]{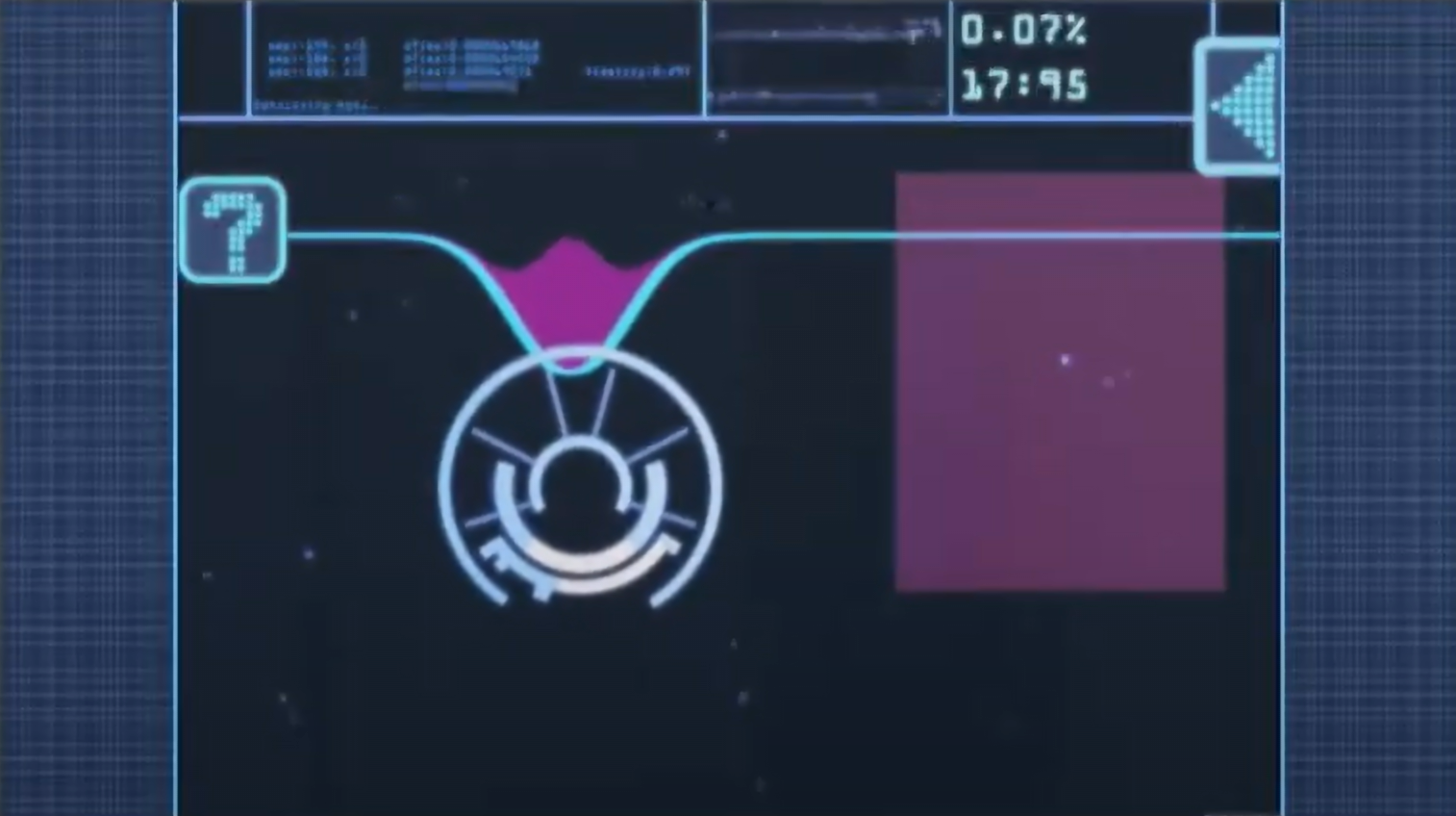}\,
\includegraphics[width = 0.3\linewidth]{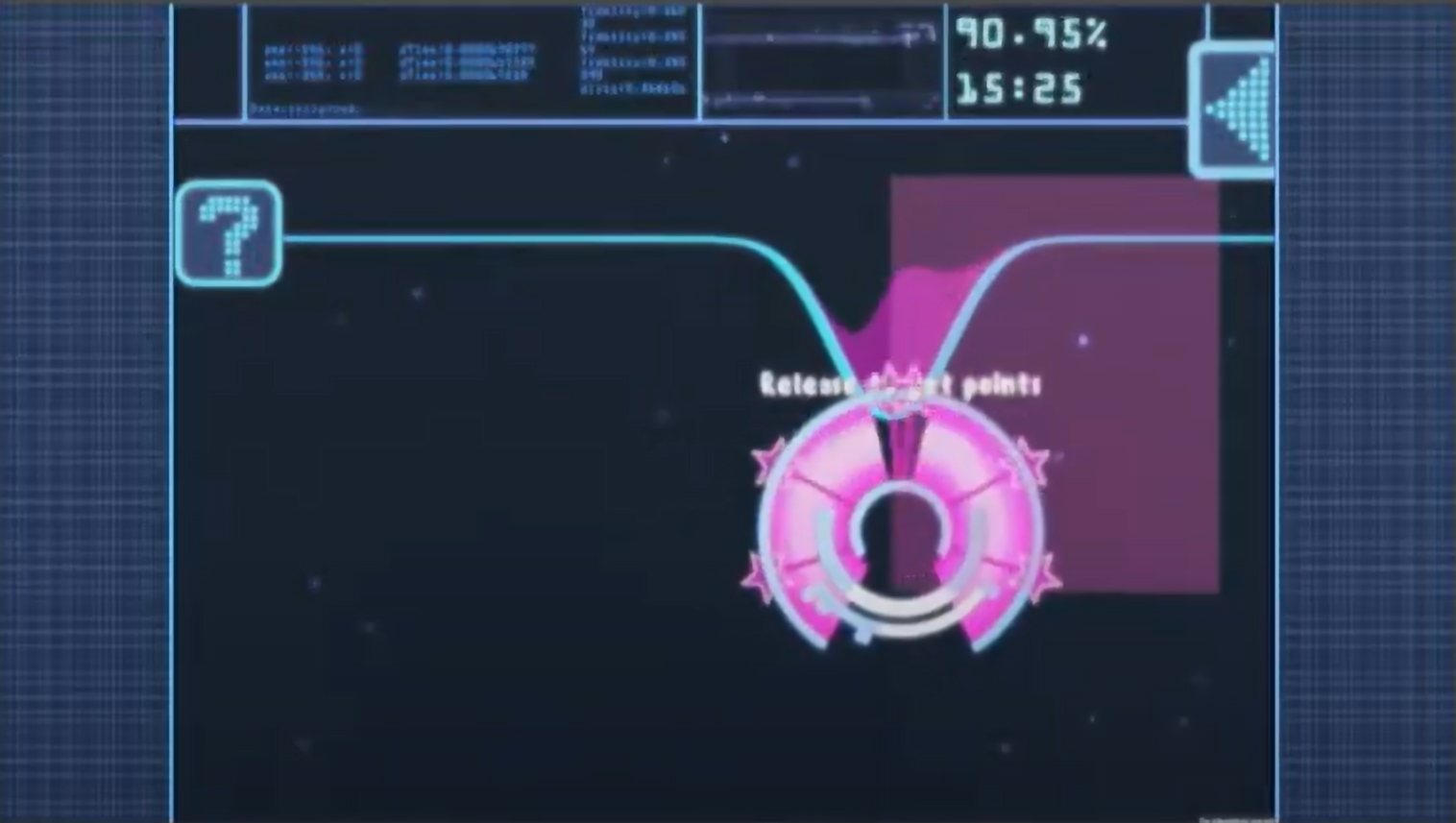}}\\
\subfloat[]{\includegraphics[width = 0.3\linewidth]{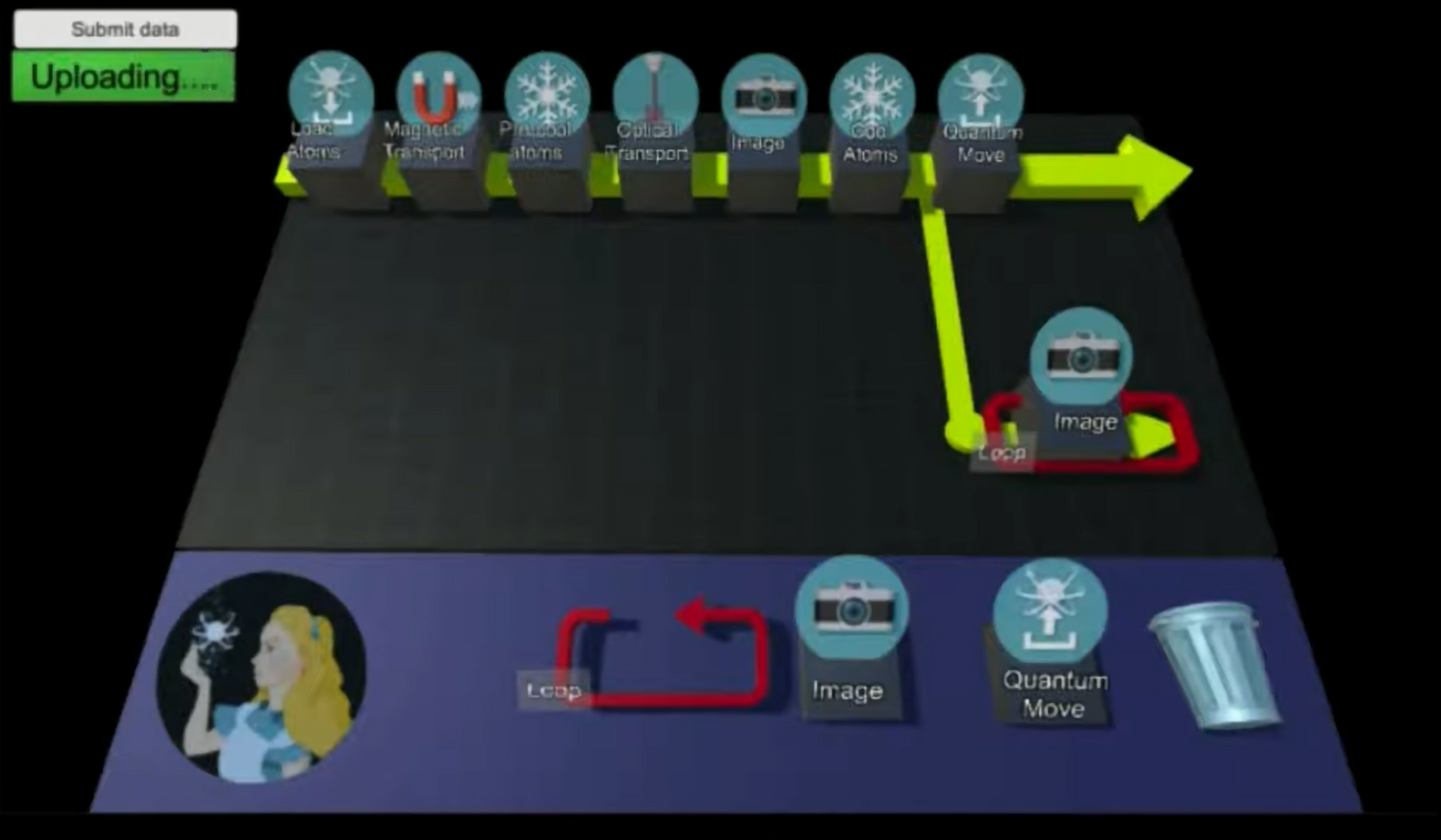}\,
\includegraphics[width = 0.3\linewidth]{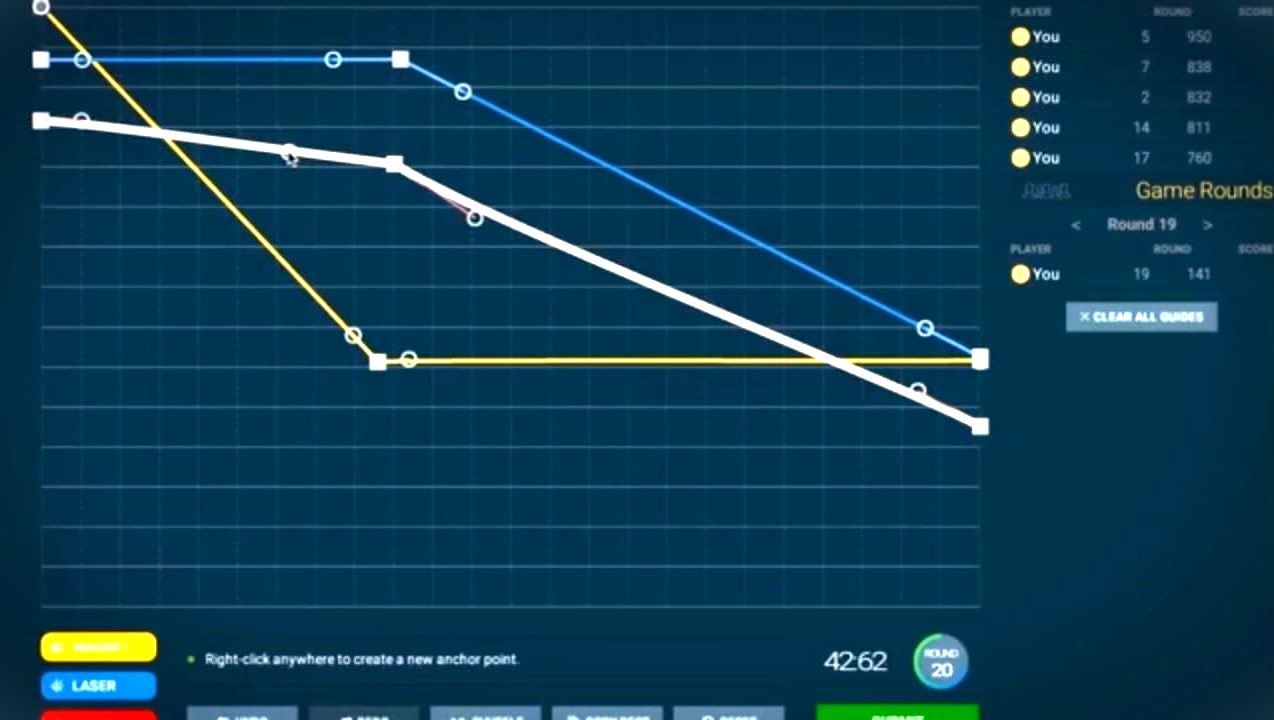}\,
\includegraphics[width = 0.3\linewidth]{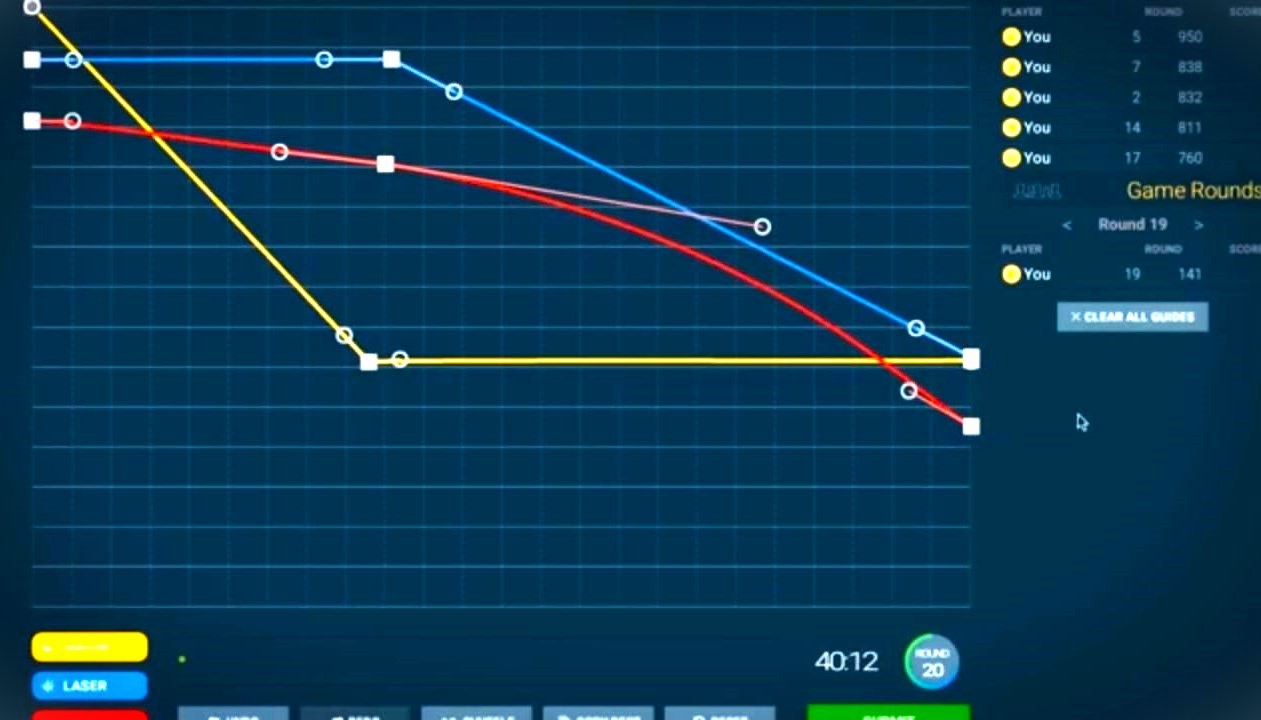}}
\caption{Screenshots from (a) the citizen science game \textit{Quantum Moves} \cite{quantummovesdemo}, (b) the online platform for \textit{Alice Challenge}. In \textit{Quantum Moves}, the player controls the position of a wave-like potential confining a liquid-like quantum object and aims to move this liquid to a specified position. \textit{Alice Challenge} allowed anyone to propose and test experimental setups through the LabView-based \textit{Alice} interface and see the actions defined by their \textit{Quantum Moves} gameplay realised with lasers.}
\label{fig:quantummoves}
\end{figure}

\subsection{First Citizen Science Games for Quantum Physics Research}
\label{cisci}
Quantum physics-related games have been developed for data gathering and problem solving in quantum sciences since the late 2010s. In 2013 the multidisciplinary research group, \textit{Science At Home}, released a beta version of the game \textit{Quantum Moves}, opened the LabView-based \textit{Alice} interface for public in 2016 on \textit{Alice Challenge}, and released the game \textit{Quantum Moves 2} in early 2019 \cite{lieberoth2014,sorensen2016,heck2018, jensen2021}. These projects are citizen science games and an interface, where players find solutions meant to optimise a certain quantum state-transfer process within the framework of quantum optimal control \cite{brif2010, koch2016}. 
\begin{figure}[ht]
\center
\subfloat[]{\includegraphics[width =0.3\linewidth]{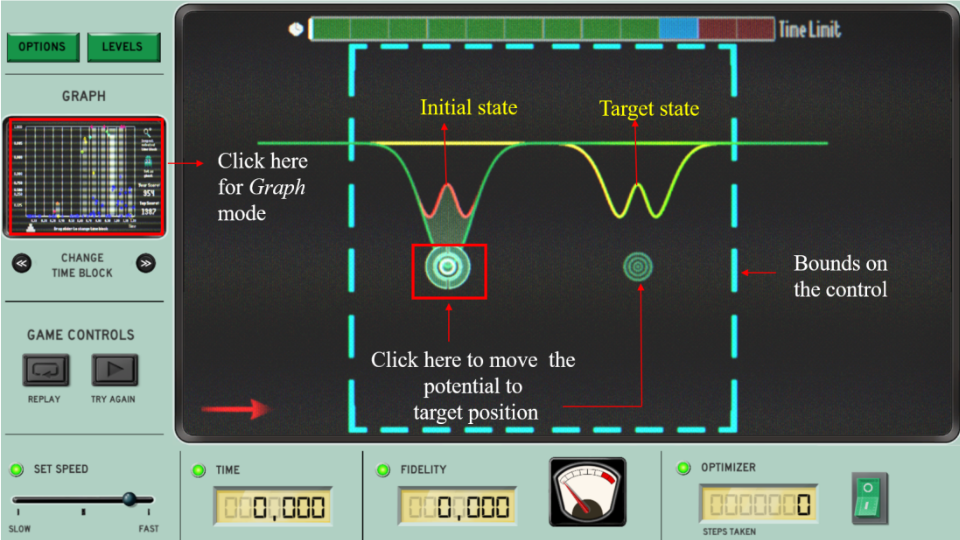}}
\subfloat[]{
\includegraphics[width = 0.3\linewidth]{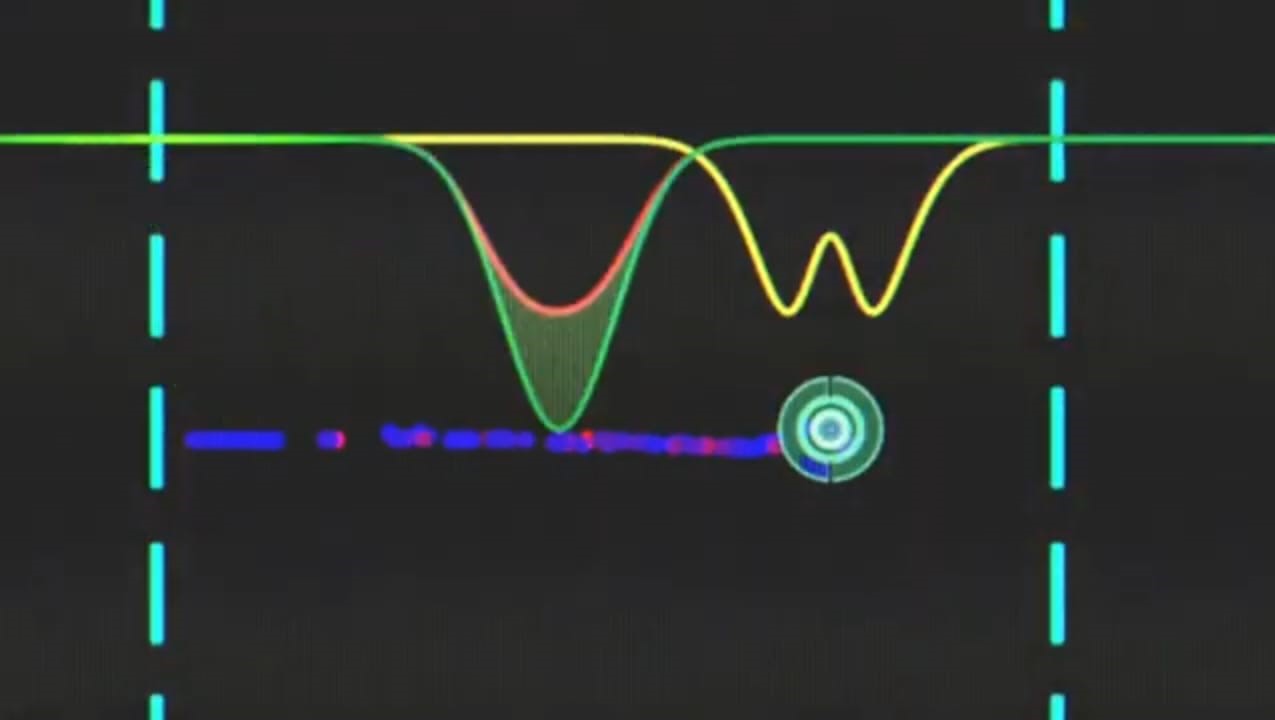}\,
\includegraphics[width = 0.3\linewidth]{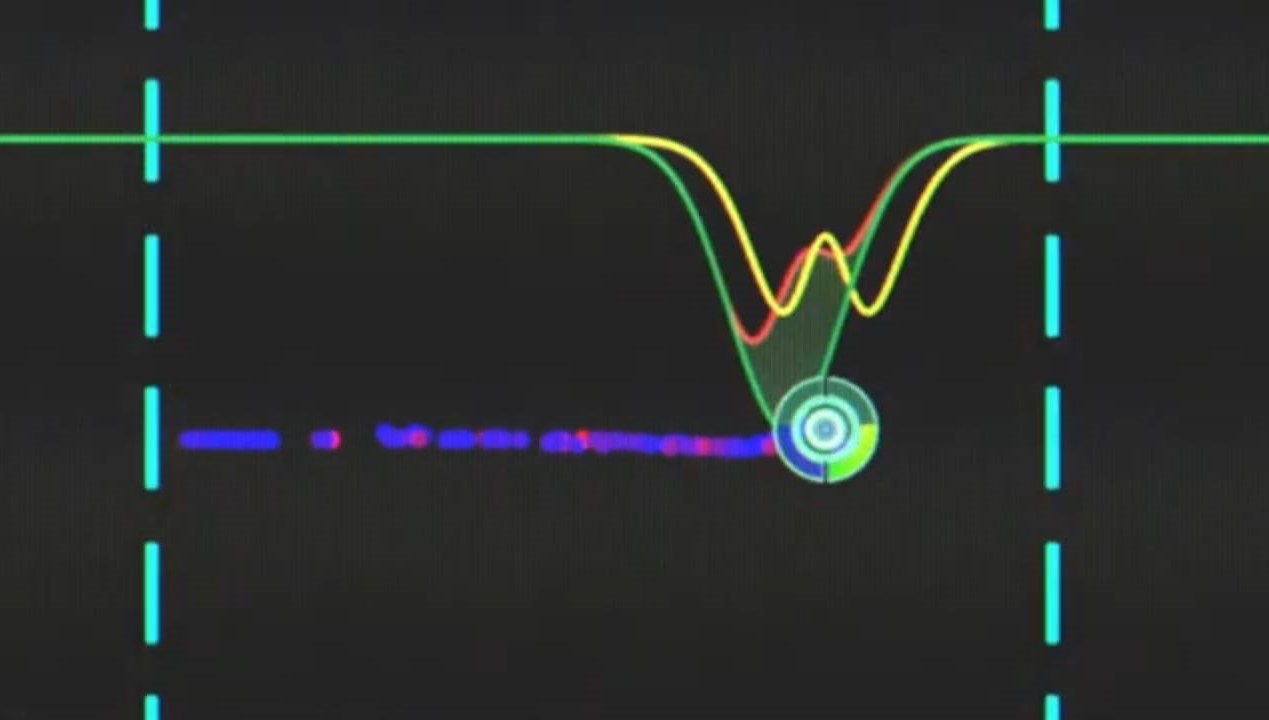}}
\caption{(a) An annotated interface of the citizen science game \textit{Quantum Moves 2} \cite{ahmedPHD} and (b) two sequential close-ups from the plot. In \textit{Quantum Moves 2}, the player controls the position of a wave-like potential confining a liquid-like quantum object and aims to move this liquid to a specified position.}
\label{fig:quantummoves2}
\end{figure}
In \textit{Quantum Moves} and \textit{Quantum Moves 2}, the player controls a 2-dimensional graph with a well-shaped confinement inside which the visualisation of a numerical quantum simulation sloshes accordingly in a water-like manner (see Figures \ref{fig:quantummoves}(a) and \ref{fig:quantummoves2}). The aim of the game is to position this “quantum liquid” at a designated spot on a depicted slope. The slope presents the external potential produced with ultra-focused laser beams and the “liquid" is a numerical simulation of the probability distribution of an ultra-cold quantum system within a crystal structure. The player controls the laser profile. When they move the object up and down they control the power or the intensity of the beam and by moving it along the left-right axis they control the precision of the beam. In \textit{Quantum Moves} this system presented a single atom, but in \textit{Quantum Moves 2} a Bose-Einstein condensate of a group of 700 Rubidium atoms was represented. The latter also provided more functionalities, including a “ghost feature" that allowed the players to be inspired by optimiser solutions.

The idea of presenting the system of lasers and the atom as a slushy liquid within an open container was applied in addition to various other citizen science game prototypes \textit{Science At Home} developed \cite{scienceathomegames, scienceathomedemo}. In \textit{Alice Challenge} the problem visualisation was “entirely abstract", as the users were to adjust three different curves that would correspond to the magnitude of a magnetic field and two laser fields (see Figure \ref{fig:quantummoves}(b)) \cite{heck2018}. The objective was to design experimental setups and study optimal ways to cool down systems consisting of Rubidium atoms through crowdsourcing. The \textit{Alice} interface allows designing setups using provided `blocks' representing lab equipment and letting players see how their moves on \textit{Quantum Moves} were run on a setup (see Figure \ref{fig:quantummoves}). 

From these approaches the players were reported reaching desired experimental solutions, the researchers were offered with valuable insight about the nature of their research problems, the observations guided the interface design in later projects, and the players were providing data for testing machine learning algorithms \cite{heck2018, jensen2021,ahmedPHD}. Yet, no further development on these games has been reported since 2021 \cite{scienceathomegames,scienceathomedocumentary}. Still, \textit{Science At Home} has since grown into a vibrant multidisciplinary hub of quantum physics researchers, game developers, psychologists and cognitive scientists who have developed educational games and citizen science games also in fields other than the study of quantum technologies \cite{ahmed2022,ahmedPHD,scienceathomegames,scienceathomedocumentary}.\\

\begin{figure}[ht]
\center
\includegraphics[width = 0.3\linewidth]{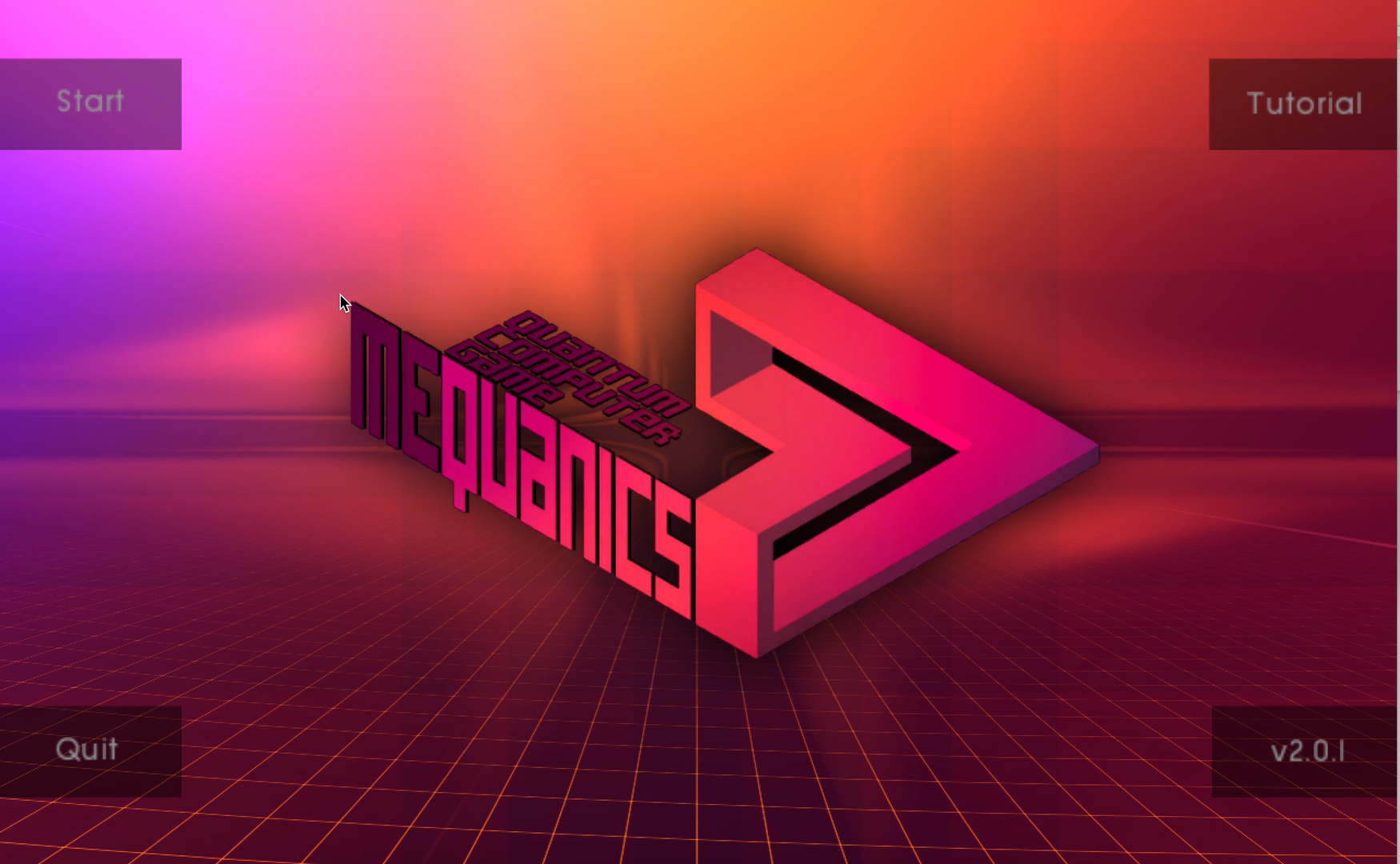}\,
\includegraphics[width = 0.3\linewidth]{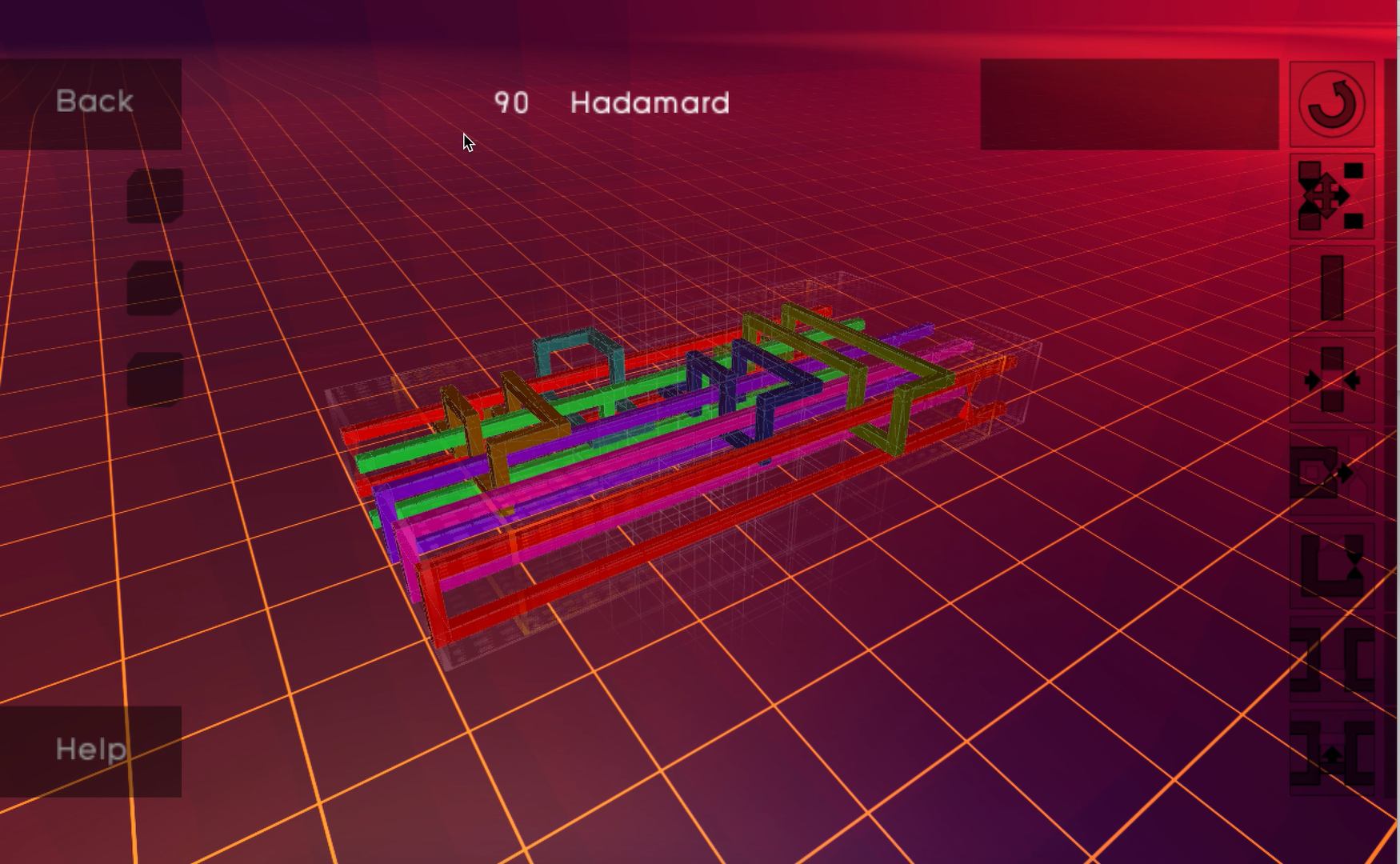}\,
\includegraphics[width = 0.3\linewidth]{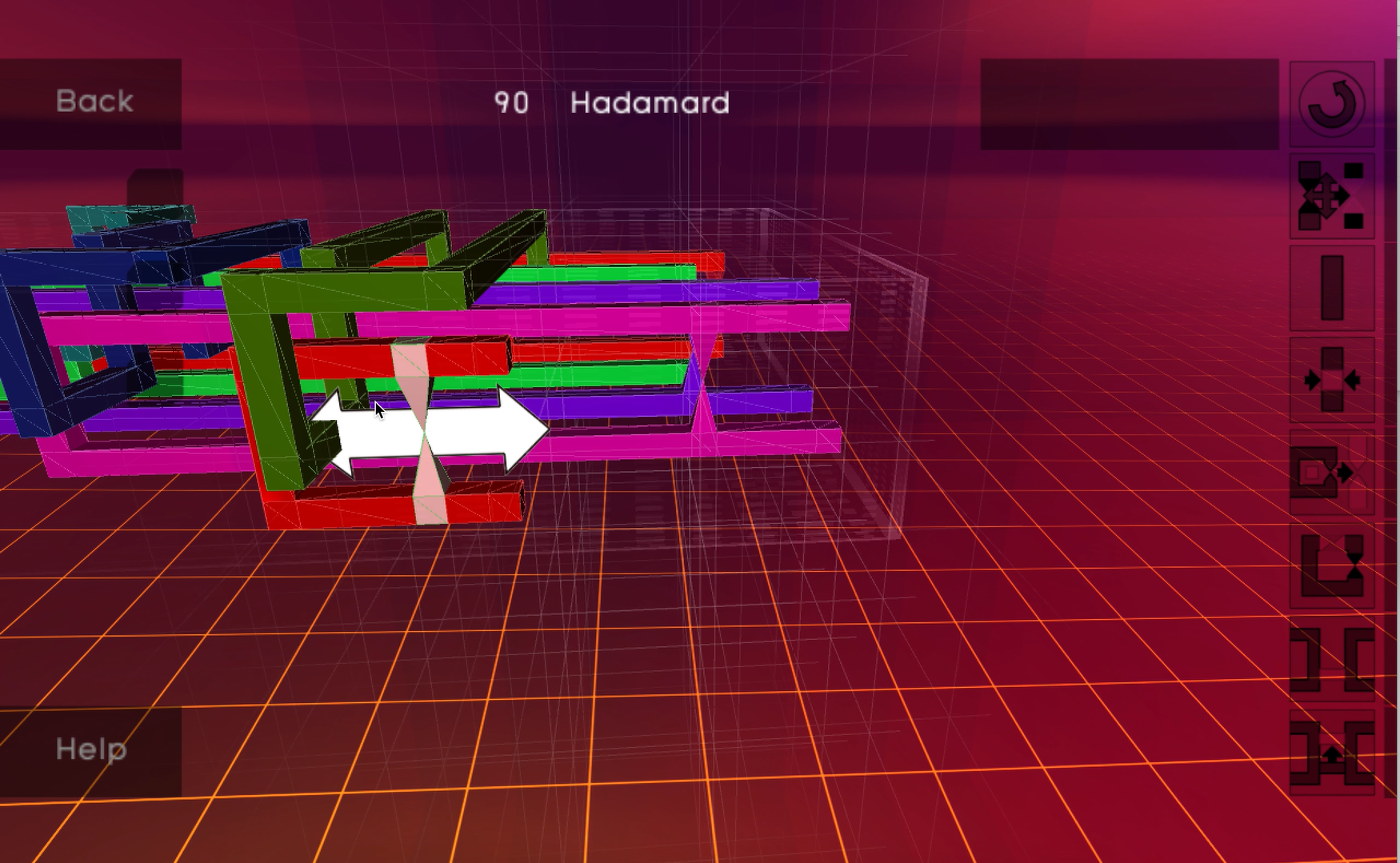}
\caption{Screenshots from the citizen science game \textit{meQuanics}. In \textit{meQuanics}, the player solves puzzles consisting of complex knot-like structures that represent topological quantum circuits of certain quantum algorithms \cite{mequanics}.}
\label{fig:mequanics}
\end{figure}
Published in 2013 by a Japanese research group working on optical quantum computers, \textit{meQuanics} was a citizen science game designed for optimising quantum error correction algorithms \cite{devitt2013,devitt2016}. The topological nature of the design of these machines allows defining a geometric structure for the execution of \textit{quantum circuits}, the computational model used in quantum computing. In the game the player aims at simplifying complex knot-like structures and minimising their size according to the rules given in the game (see Figure \ref{fig:mequanics}). This procedure corresponds to an optimising routine for the algorithm. These processes from the players were aimed to be used to train a machine learning algorithm for solving the optimisation of these error correction algorithms. Designed for touch-based platforms and online collaboration, the game claims to be the first crowd-sourced project related to quantum computing attempting to use the \textit{creativity} of the human mind to provide usable data for the development of quantum hardware. \textit{meQuanics} is still openly available for exploring a set of quantum algorithms in the form of topological quantum circuits, though the game has been reported as being at a prototype level with no recent updates \cite{devitt2016,mequanics}.

\begin{figure}[ht]
\center
\includegraphics[width = 0.24\linewidth]{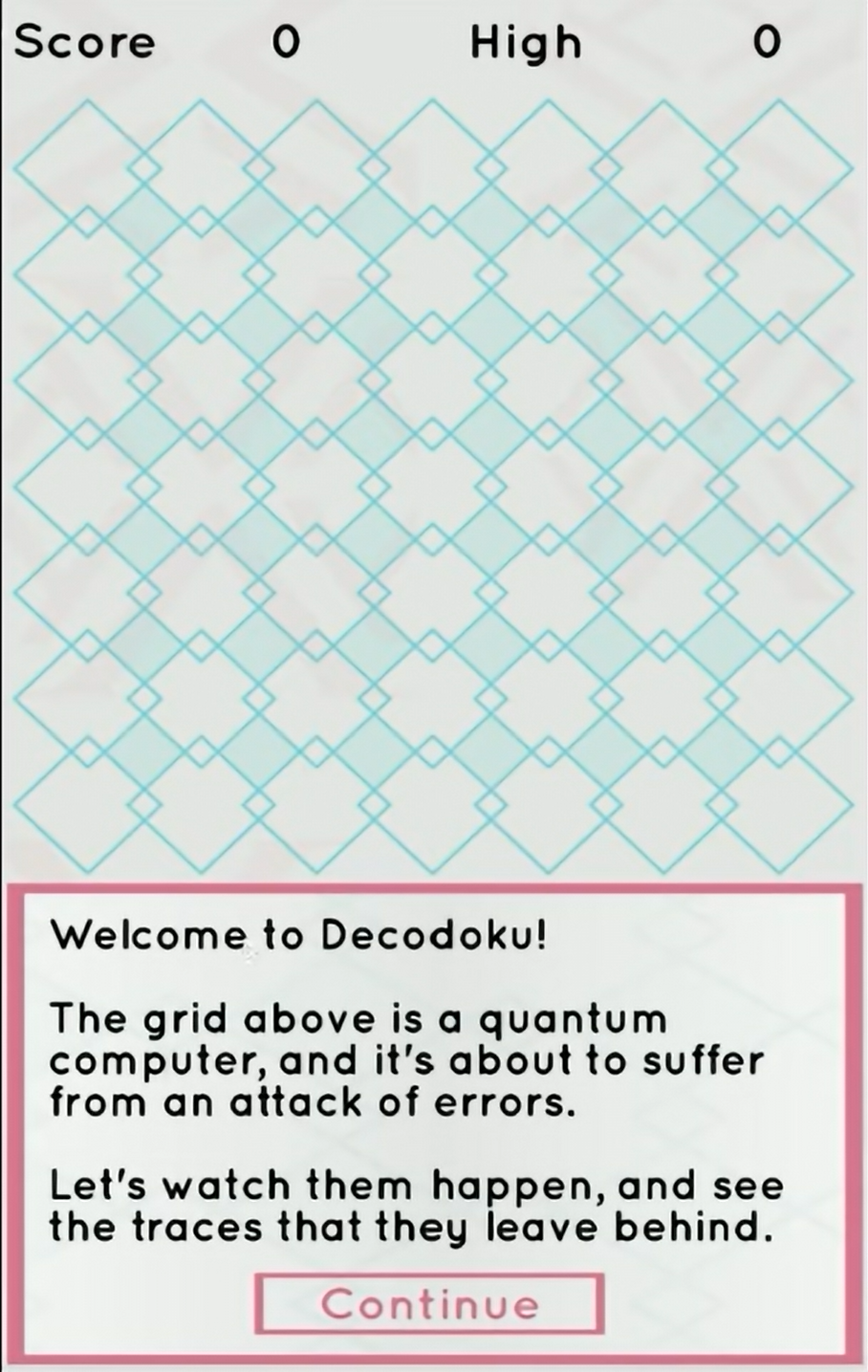}\,
\includegraphics[width = 0.24\linewidth]{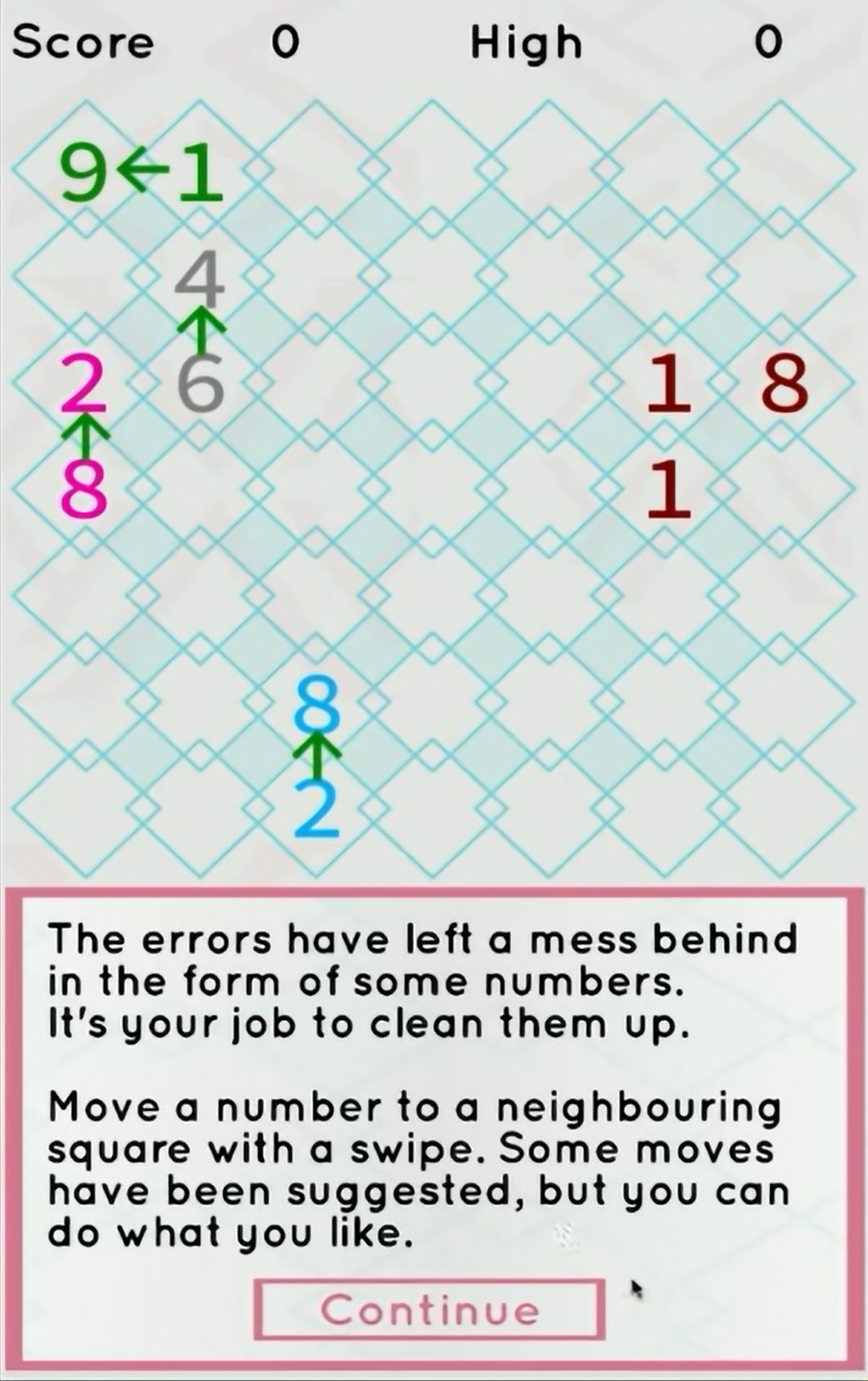}\,
\includegraphics[width = 0.24\linewidth]{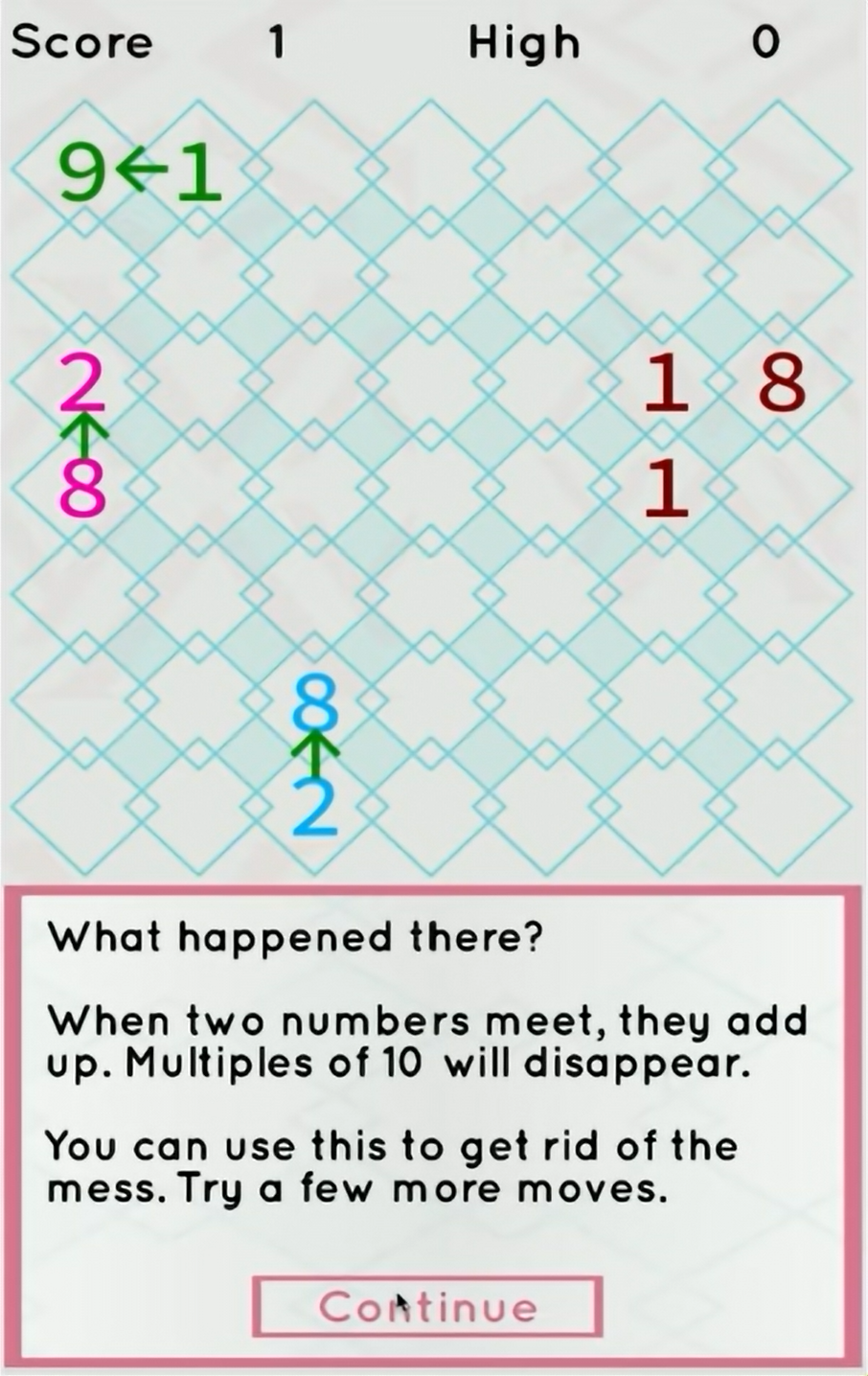}\,
\includegraphics[width = 0.24\linewidth]{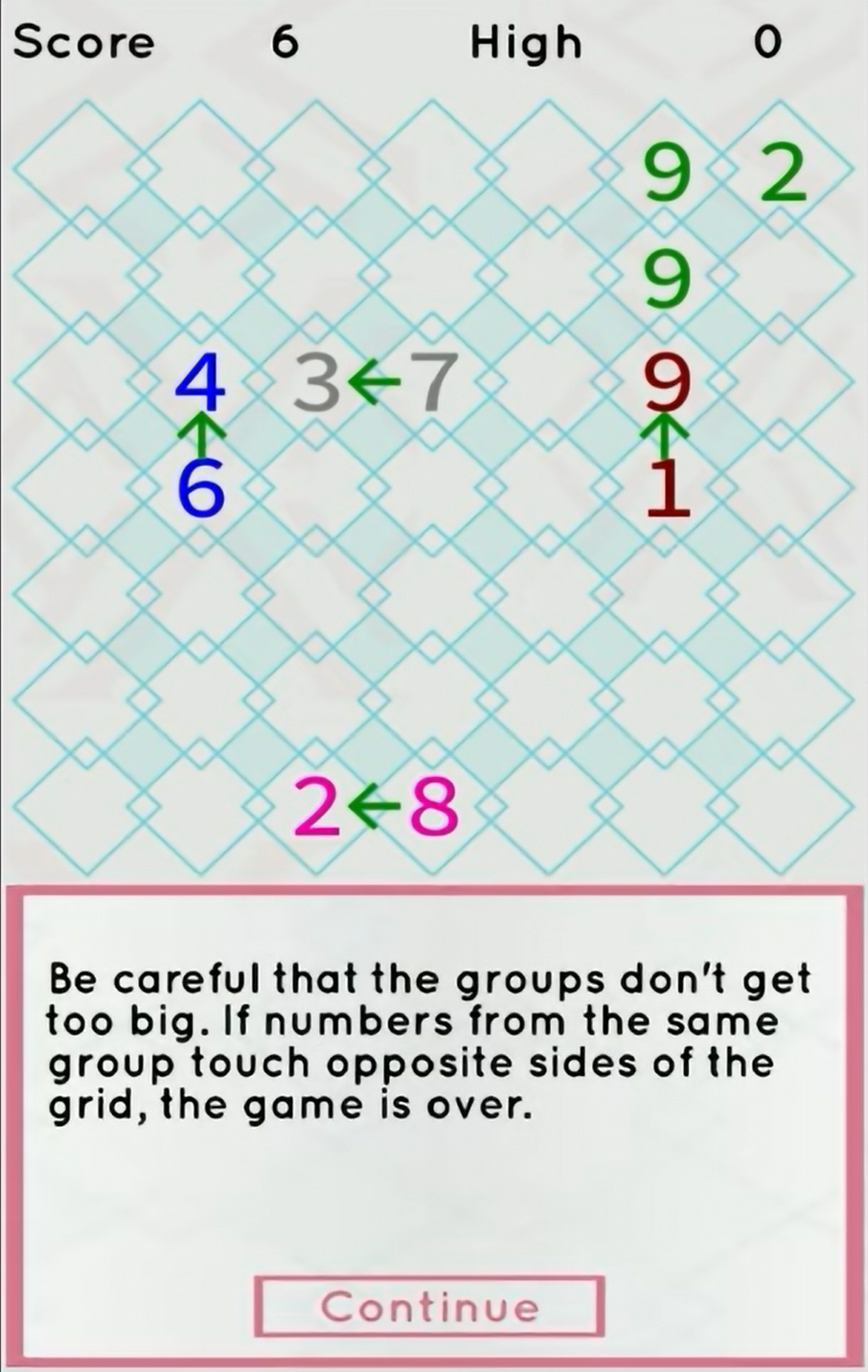}\,
\caption{Sequential screenshots taken from the tutorial phase of the game \textit{Decodoku}, a puzzle game for developing error correction protocols \cite{wootton2017decodoku}.}
\label{fig:decodoku}
\end{figure}
Opened for public involvement in January 2016, the citizen science game \textit{Decodoku} was a puzzle game on error correction protocols for quantum computers that aimed to share and demystify the work of scientists and break preconceptions about the nature of quantum physics \cite{wootton2017}. Instead of the conventional 2-level qubit, \textit{Decodoku} concentrated on the multi-level computational unit alternatives, \textit{qudits} \cite{nielsen2002}. As the already-known decoding methods for them had been of a heuristic nature, the idea was to develop topological quantum error-correcting algorithms based on the methods that players would use to solve puzzles. The game had three versions and a tutorial mode for educating about quantum error correction without the requirement of prior mathematical knowledge (see Figure \ref{fig:decodoku}). The game did not gather data on the moves of the players during the gameplay, but left the task of analysing the solution methods to the players, who were able to discuss them on an open forum \cite{wootton2017decodokureddit}. This method led to several highly satisfactory results and the game was awarded at the \textit{qstarter awards} by the The National Center of Competence in Research `Quantum Science and Technology' in Switzerland \cite{wootton2017,qsit}. \\

On November 30th 2016, a game called \textit{The BIG Bell Quest} was deployed for creating human-generated randomness for testing the fundamental theory of quantum information, the Bell test \cite{bell1964,bellgame2018}. In the game, the players made sequential choices regarding two options for a path that the character follows, which then generated bits directed to 13 separate experimental groups with various physical tests. The player activity was synchronised with experimental operations. 100,000 participants around the world created over 97 million binary choices that were used to determine measurement settings in 13 laboratories across five continents. 

This experimental setting was inspired by an idea originally by John Bell, who proposed that human choices, relying on free will, could provide truly unpredictable free variables for his test, the Bell test \cite[Chapt. 7]{bell1987}. Prior attempts had relied on machine-reliant randomness and on cosmic sources, which were not always seen as reliable against the claim of pre-set “hidden variables". Sourcing human-generated randomness for a large enough setup and for creating enough choices for a statistically significant test required the modern tools of crowdsourcing and gamification. Human-generated randomness was used in the worldwide experiment to ensure that the experimental setup had no open loops regarding randomness, and would therefore prove that quantum entanglement does indeed challenge the notion of local realism. 

\subsection{Games on Quantum Computers and Simulators} 
\label{qcomp}
\textit{Quantum computers} are computers whose computing power relies on superposition, entanglement, and other quantum physical phenomena. They were proposed in the early 1980's and the first physical realisations have been available for public access for the last ten years \cite{benioff1980,feynman1982,deutsch1985, divincenzo1998, divincenzo2000, nielsen2002, ladd2010}. The basic units for a quantum computer are quantum bits, otherwise known as \textit{qubits}. Qubits work in a fundamentally different way than bits as, instead of the classical “zero-state" and “one-state", qubits are able to exhibit a quantum superposition of both. In a quantum computer, it is possible to manipulate and control their quantum states and take advantage of properties like entanglement in computational processes. This peculiarity of quantum computing is the reason behind claims of these devices bringing unparalleled prospects particularly in finance, drug development, industrial optimisation problems, molecular biology, and cryptography and has therefore gained attention from both media and industry \cite{quantumalgorithmzoo, shor1994,horowitz2019,outeiral2021,fox2020}. Once feasible, scalable and reliable, quantum computers are expected to offer a considerable advantage over even the most prominent supercomputers when it comes to specific problems like optimising a route with several stops, efficient database searches, and finding the correct structures of proteins \cite{marsh2020, grover1996, robert2021}.\\

The first games on quantum computers were made in 2017 on the first publicly accessible IBM Quantum devices as command-line based adaptations of \textit{Rock, Paper, Scissors} and \textit{Battleship} \cite{wootton2018history, wootton2018GameOn,piispanen2023history}. Particularly since the release of \textit{Qiskit}, an open-source software development kit by IBM for quantum computing, experimental realisations of quantum game theoretic scenarios have also been presented \cite{du2002,khan2018,szabo2023}. Quantum game theory is the discipline expanding game theory to include quantum mechanical phenomena \cite{meyer1999, khan2018}. Quantum game theory deepens the understanding of quantum information processing and quantum algorithms and the division between the quantum and classical worlds. 

\begin{figure}[ht]
\center
\subfloat[]{\includegraphics[width = 0.3\linewidth]{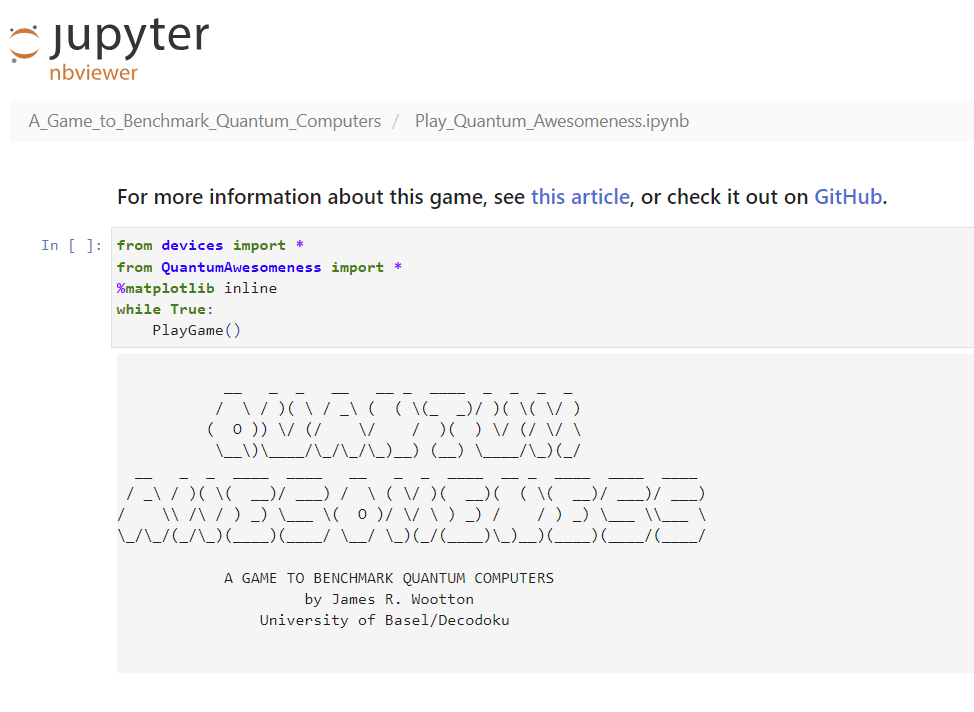}}\,
\subfloat[]{\includegraphics[width = 0.3\linewidth]{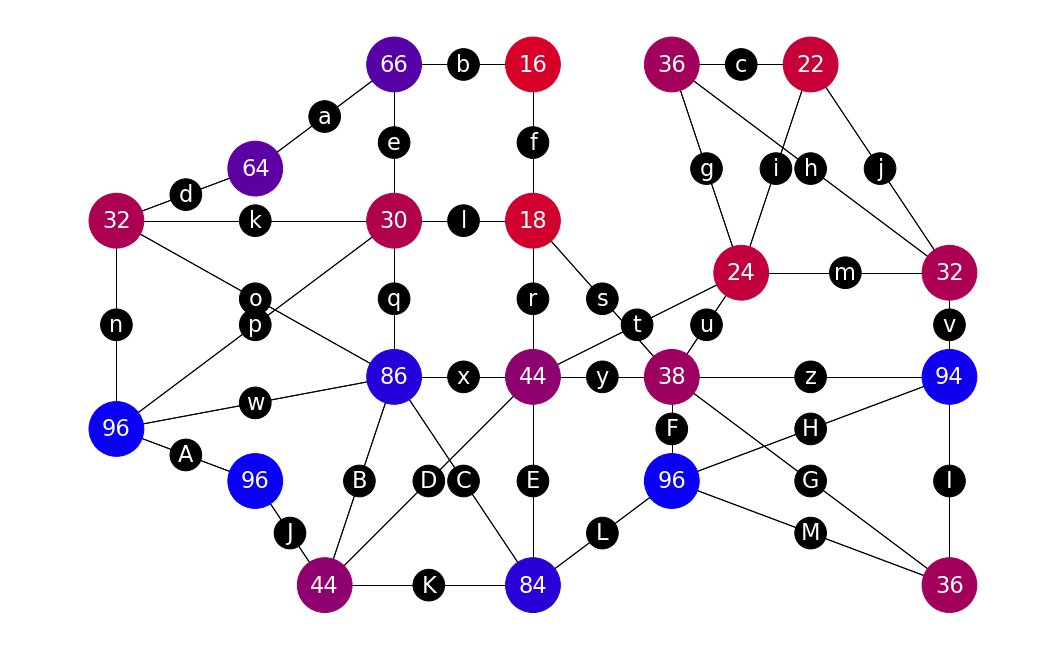}}\,
\subfloat[]{\includegraphics[width = 0.3\linewidth]{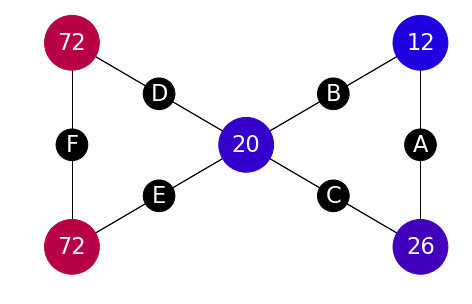}}\,
\caption{(a) A screenshot from the Jupyter interface and example puzzles from the game \textit{Quantum Awesomeness}, showing the structural depiction of two IBM Quantum devices: (b) the 20 qubit \textit{QS1\_ 1} device and (c) 5 qubit \textit{ibmqx2}. \textit{Quantum Awesomeness} offers a series of puzzles designed for benchmarking quantum computers. Each puzzle is a connectivity graph made up of a grid of qubits presented as coloured dots with numbers in them and paired with letters (in black, smaller circles). The player pairs the coloured circles so that each dot should have a similar colour and number as one of their neighbours.}
\label{fig:screenshotsqc}
\end{figure}
For the use of citizen science, not many realisations have existed that run \textit{on} a quantum computer. The puzzle game \textit{Quantum Awesomeness} was designed to find the path towards “quantum computational supremacy" (see Figure \ref{fig:screenshotsqc}) \cite{wootton2018}. While initially conceived for educational purposes to provide players with insights into the capabilities of a quantum device, \textit{Quantum Awesomeness} was proposed also as a benchmarking tool for the early quantum computers with citizen scientists providing suitable quantum programs, \textit{quantum circuits}, through the gameplay \cite{wootton2018, wootton2018medium, wootton2020}. By attempting to solve puzzles in the game, mistakes made by players contribute to the underlying quantum program, which is useful for \textit{random circuit sampling}, and has been suggested for benchmarking quantum devices of the time \cite{boixo2018characterizing}. Producers of quantum computers were encouraged to run the program on their quantum machinery and share the connectivity graphs of them for the game to use. The game could have this way been used as a citizen science sourced standardised way of comparing early quantum hardware.  

The game was first developed as command line or Jupyter interface playable (see Figure \ref{fig:screenshotsqc}(a)) \cite{woottonawe}. Yet no specific problems suitable for citizen science engagement were identified during its development. According to the author of the game\footnote{Reference to personal email exchange with James R. Wootton. Permission to cite granted in the discussion.}, there remains a prospective opportunity for the revival of \textit{Quantum Awesomeness} in the future, especially considering its conceptual alignment with new layer fidelity benchmark \cite{wack2023}. However, the viability of this notion is currently uncertain, and further exploration is required to ascertain its potential impact within the realm of citizen science.

\begin{figure}[ht]
\center
\subfloat[]{\includegraphics[width = 0.3\linewidth]{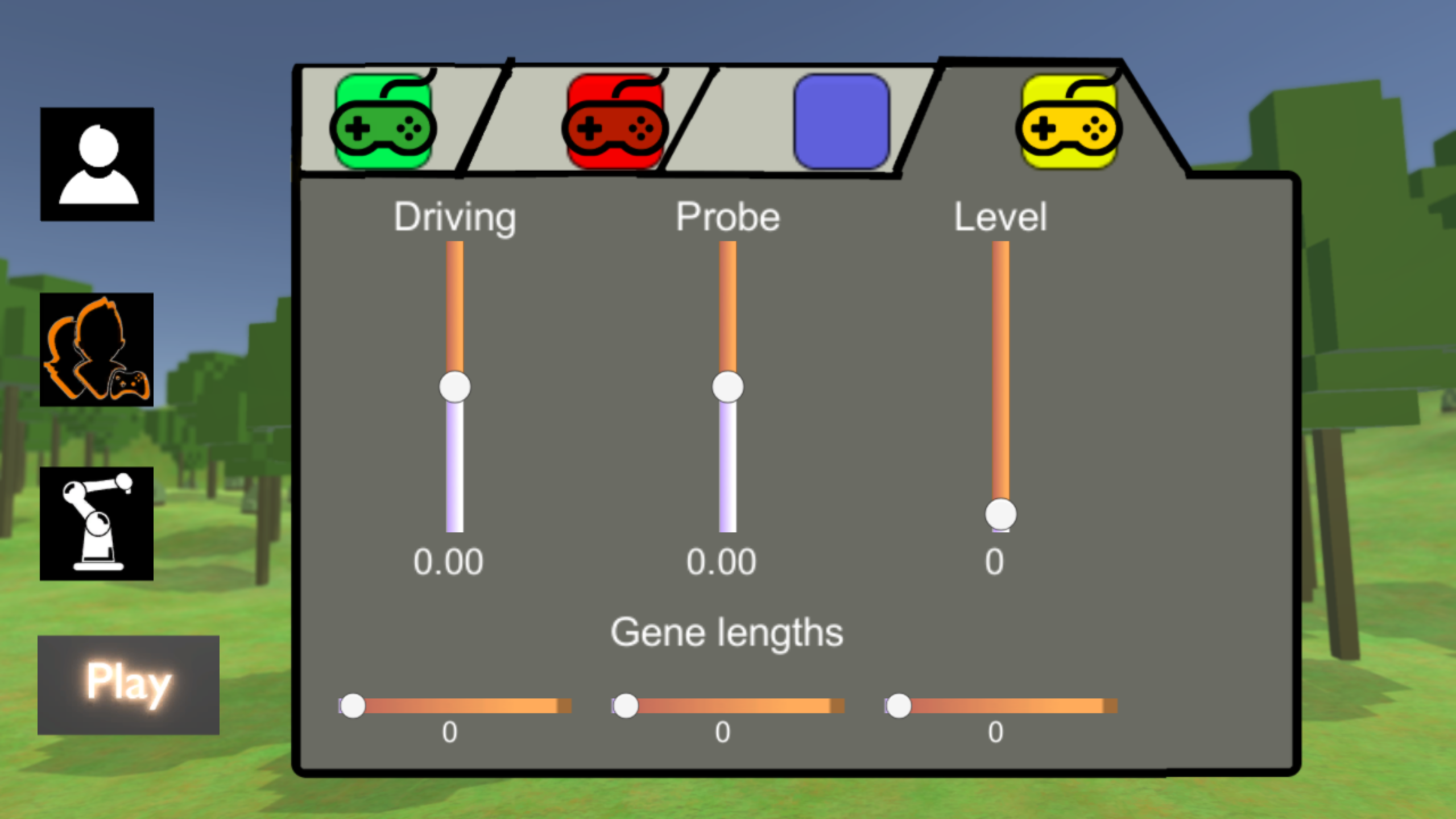}\,
\includegraphics[width=0.3\linewidth]{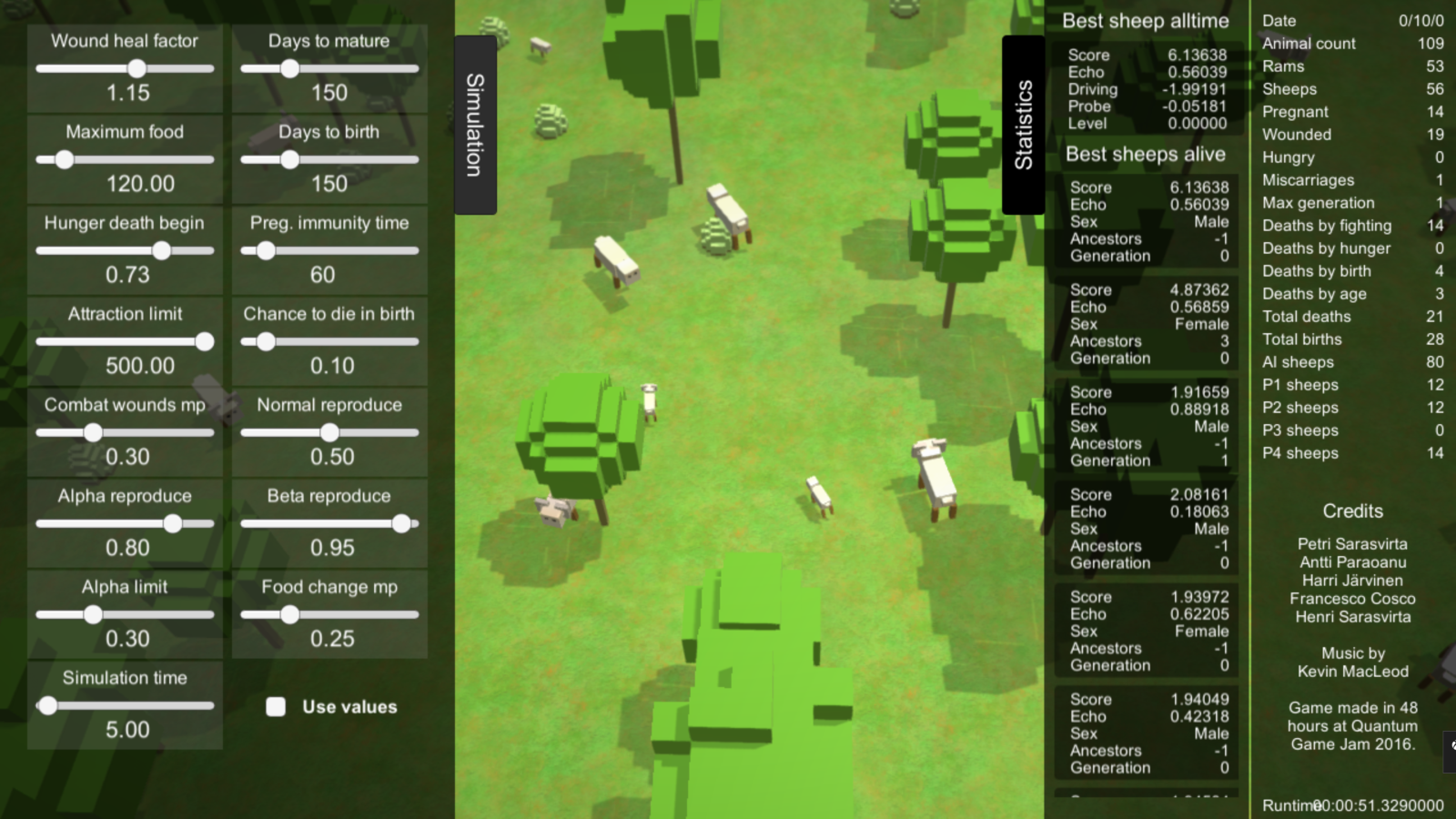}\,
\includegraphics[width=0.3\linewidth]{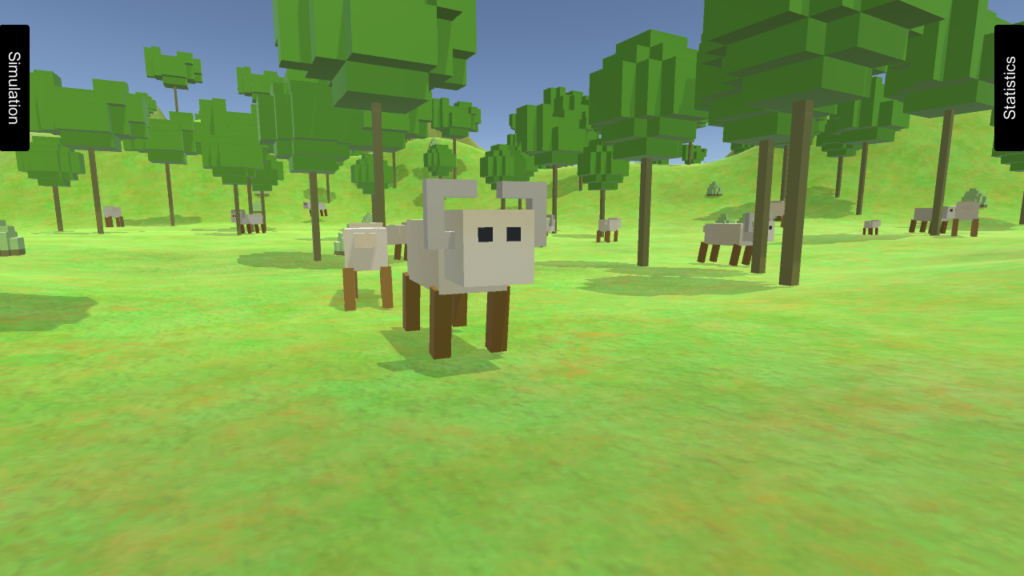}}\,
\subfloat[]{\includegraphics[width= 0.3\linewidth]{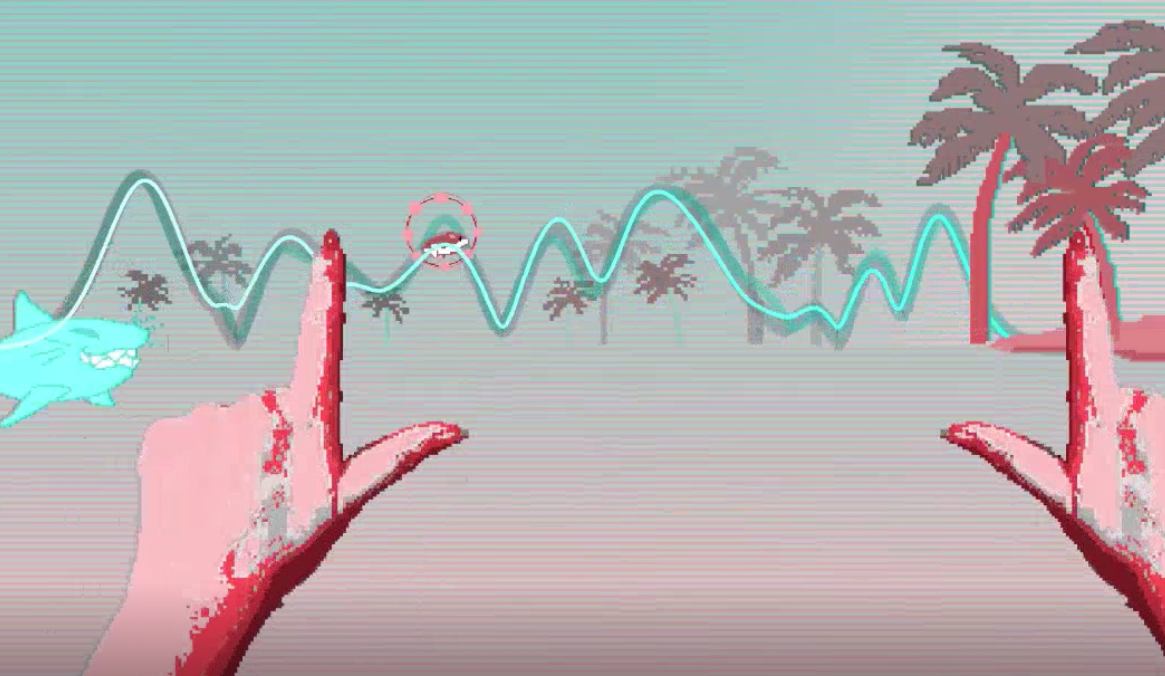}}\,
\subfloat[]{
\includegraphics[width=0.3\linewidth]{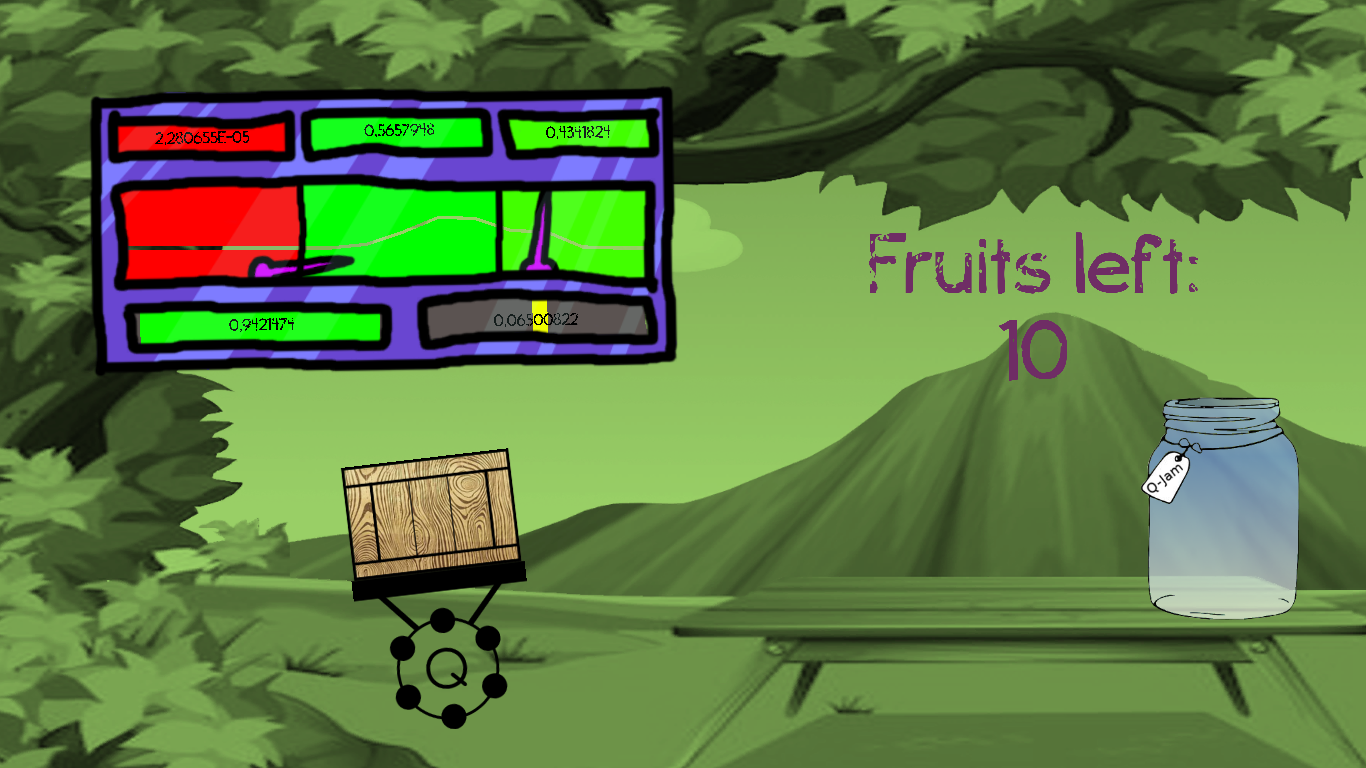}\,
\includegraphics[width=0.3\linewidth]{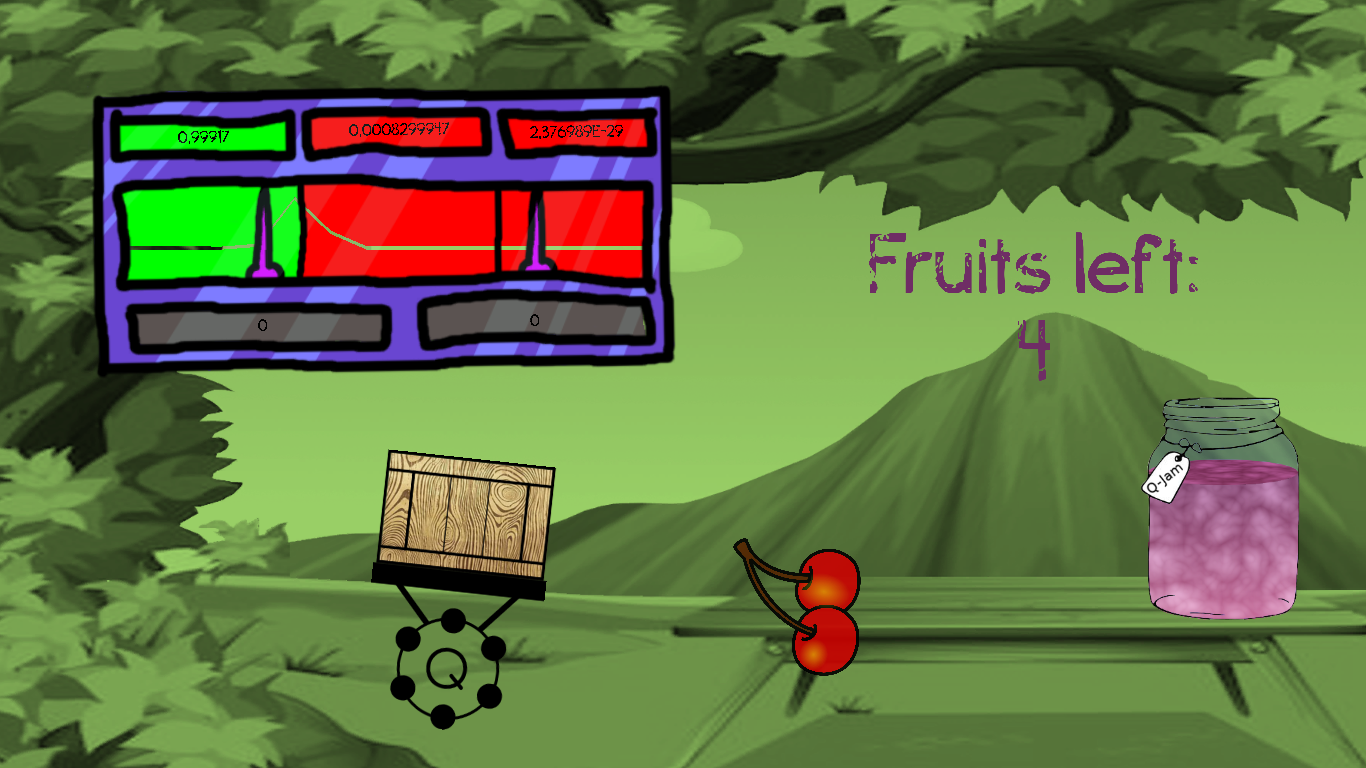}}\,
\caption{Screenshots from games (a) \textit{Quantum Sheep}, (b) \textit{Hamsterwave} and (c) \textit{Quantum Fruit}. \textit{Quantum Sheep} is a simulation of a flock of sheep developed during the Quantum Game Jam in 2016. \textit{Hamsterwave} and \textit{Quantum Fruit} were developed during the Quantum Game Jam in 2019 called \textit{Quantum Wheel} \cite{qwheel}. In \textit{Hamsterwave}, the player controls the shape of the wave to save a hamster. In \textit{Quantum Fruit}, the player aims to hit a glass jar with fruit by controlling a throwing device. The control of the multiplication of the sheep, the shape of the wave in \textit{Hamsterwave}, and the behaviour of the throwing device in \textit{Quantum Fruit} stem from a numerical simulation developed for citizen science game prototypes \cite{piispanen2024}.} 
\label{fig:screenshots02}
\end{figure}

\subsection{The Rise of Public Quantum Game Development}
\label{gamejam}
Not all quantum games have been developed by researchers or research groups \cite{kultima2021qgj, piispanen2023history, piispanen2022}. \textit{Quantum Game Jam}s are weekend-long events, where quantum physicists join forces with game developers, artists, and hobby game creators to produce quantum games or other playful creations according to a common theme related to quantum physics \cite{kultima2021qgj,piispanen2023qgj}. These annual events started as a series of six hybrid events between the years 2014 and 2019, where citizen science played a role in the themes. Game jams and hackathons have been found to be fruitful for serious game prototyping \cite{ramzan2016,abbott2023}. The event organisers of Quantum Game Jam were motivated to inform the wider public about the ongoing advancements within the research of quantum technologies, but also hoped to provide an inspiring platform for the creation of citizen science game prototypes for research in quantum optimal control theory \cite{mequanicssab,kultima2021qgj}. For supporting the development of citizen science game prototypes, a simulation tool called the \textit{Quantum Black Box} (QBB), aimed to be incorporated into citizen science game prototypes, was developed \cite{piispanen2023projects, piispanen2024,qbb}. As an example, in 2016, the game \textit{Quantum Sheep} incorporated this numerical simulation of a simplified Bose-Einstein condensate. The simulation was connected to the virtual evolution of a flock of sheep the player was to foster (see Figure \ref{fig:screenshots02}(a)). The QBB simulator was further developed into an interface that was first implementable with the developed games as a locally running Python server and later also as a Unity package containing a wrapper for the Python package \cite{piispanen2024}. In 2019, this QBB package also included a demo game aimed to give a head start into the citizen science game development by showing how one can interact with the code \cite{kultima2021qgj, piispanen2024}.

\begin{figure}[ht]
\center
\subfloat[]{\includegraphics[height = 0.28\linewidth]{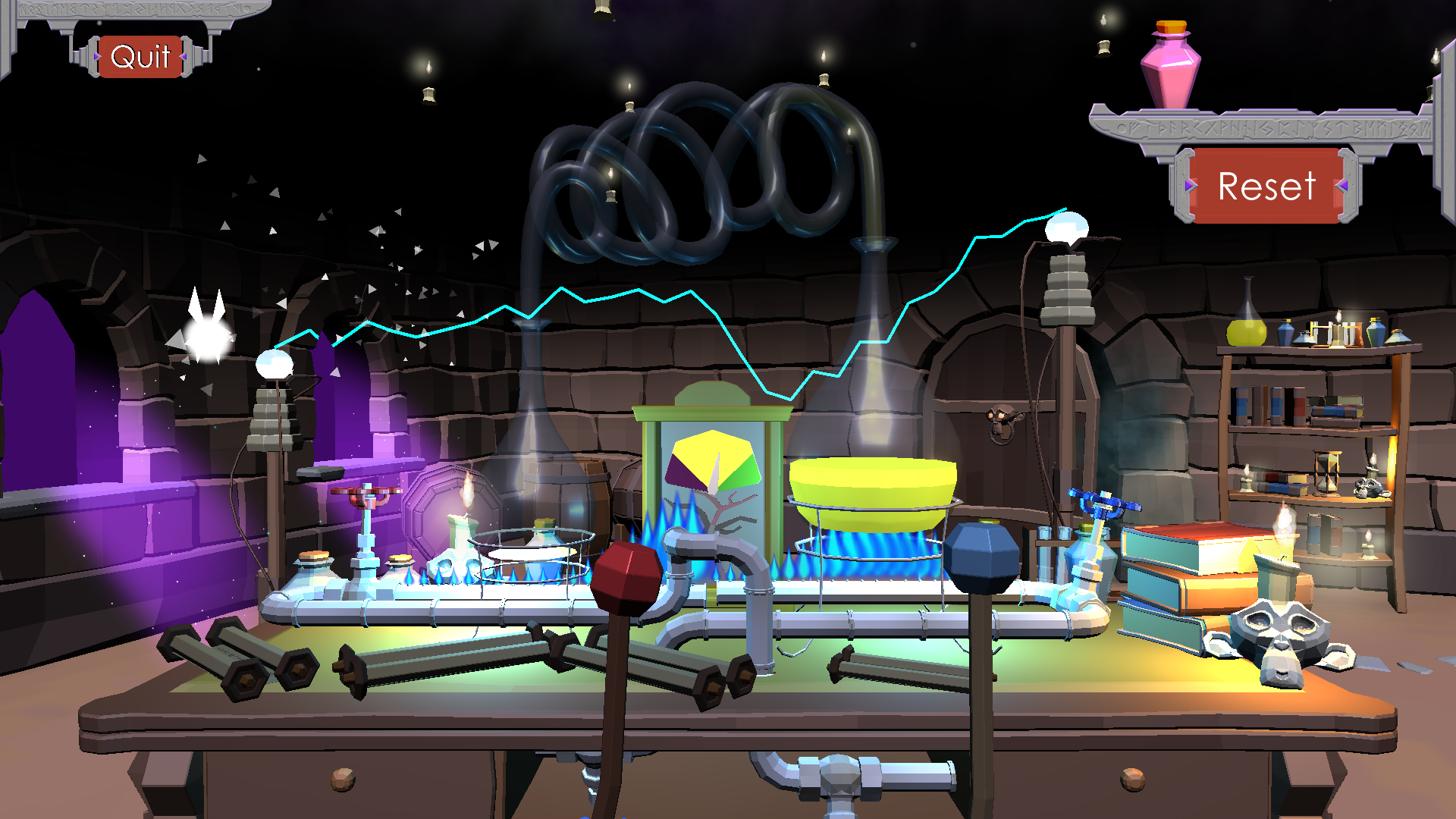}}\,
\subfloat[]{\includegraphics[height = 0.28\linewidth]{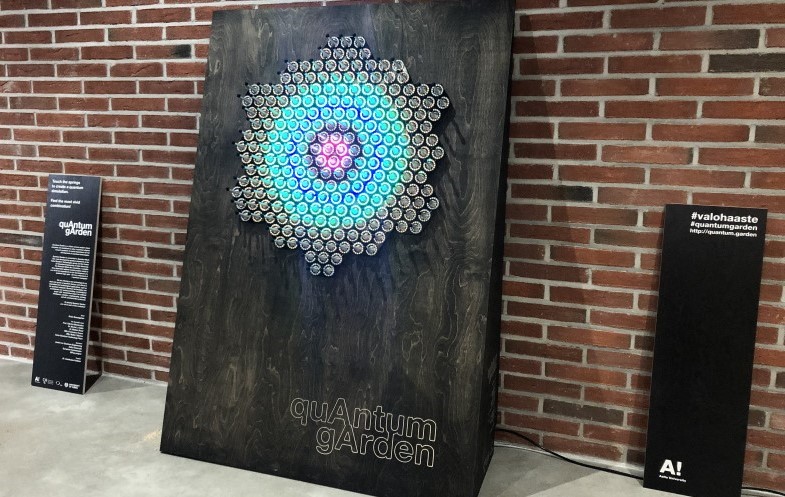}}
\caption{(a) Screenshots from the citizen science game \textit{QWiz} and (b) a picture of the \textit{Quantum Garden} installation \cite{qgarden}. \textit{QWiz} was developed as a collaboration between a game development company and the research group behind the first six \textit{Quantum Game Jam} events that aimed at fostering the creation of citizen science game prototypes \cite{piispanen2024}. \textit{Quantum Garden} is an interactive light installation. Both \textit{QWiz} and the first iteration of \textit{Quantum Garden} incorporated a numerical simulation of a simple Bose-Einstein condensate developed for citizen science game prototypes.}
\label{screenshotspiispanen}
\end{figure}
Between 2014 and 2019 Quantum Game Jam events hosted the creation of 68 games and from them 11 games made use of the QBB simulator \cite{kultima2021qgj}. Due to the 48-hour restriction on developing the games, some of these games used the QBB simulator merely as a random number generator, but some thrived as highly attractive and enjoyable experiences using the simulator in a meaningful manner. One of the most attractive games was \textit{Hamsterwave}, where the shape of the probability distribution calculated by the QBB simulator represents a wave on which a little hamster sails (see Figure \ref{fig:screenshots02}(b)) \cite{qwheel}. The game was further developed to suit exhibition showcases, but failed to acquire resources for development into a citizen science game \cite{piispanen2023projects,piispanen2024}. From the games that used the provided QBB simulator as intended, the 2019 creation \textit{Quantum Fruit} allowed players to reach a known optimal solution for the underlying version of the optimisation problem, which was considered as a proof-of-concept for the QBB \cite{qwheel,piispanen2024}. In \textit{Quantum Fruit} the player controls a delicate device for throwing fruits to a jar. The trajectory of the fruit is determined by the QBB simulator and the player is offered with visual indicators of the chosen parameters (see Figure \ref{fig:screenshots02}(c)). The citizen science game prototypes developed at Quantum Game Jams served as an inspiration for the citizen science game \textit{QWiz}, where the player controls a liquid within a complicated laboratory setup (see Figure \ref{screenshotspiispanen}(a)) \cite{piispanen2024}. \textit{QWiz} used the same numerical simulation, QBB, but was developed outside the event by the research group behind the Quantum Game Jam events of 2014 - 2019 together with a professional game development company \cite{piispanen2023projects, piispanen2024}. The game \textit{QWiz} was published as a virtual reality game, a browser version and a tablet version.

In addition to quantum physics-related game prototypes, the QBB simulator developed for the Quantum Game Jams and used in the citizen science game \textit{QWiz} was employed in the commissioned artwork \textit{Quantum Garden} \cite{qgarden, piispanen2023projects, piispanen2024}. \textit{Quantum Garden} is an interactive light installation consisting of spring-like door stoppers acting as the controls of the installation, each surrounded by a set of multicoloured LED lights mounted on a wooden panel (see Figure \ref{screenshotspiispanen}(b)). On the first iteration of the installation, a touch of a spring would send a signal to the computer of the installation, where the QBB simulator would send feedback in the form of a colourful animation indicating the suitability of the initial selection. A rewarding special animation was lit when the user found a selection of the springs that would correspond to the most preferable outcomes. No recording on the possible success of this installation as a citizen science tool is presented and later iterations included a different numerical simulation not related to citizen science \cite{piispanen2024}. \\

The 2019 Quantum Game Jam was also the first-ever game development event with a dedicated access to quantum computers \cite{kultima2021qgj}. The event was organised together with IBM Research in Helsinki, Finland, and several quantum computing themed game jams and hackathons have been organised since. The Quantum Game Jam event turned online in 2021 and has continued on a yearly basis since \cite{piispanen2023qgj}. To date, there are over 180 quantum game jam games, with well over 30 quantum physics-related games from hackathons and some 20 games from university courses like the \textit{Aalto Quantum Games} course \cite{quantumgames}. Many of these games have either been developed using a simulator of the quantum computing software such as \textit{Qiskit} offered by IBM Quantum or even use actual quantum computers, but since 2019 these games have not been reported being developed with citizen science in their design motivations. 

\section{Discussion}
In this article we have learned about eleven citizen science quantum games. In addition we viewed one citizen science platform designed for crowdsourcing ideas for experimental setups and an interactive art piece. 
The available literature and resources, particularly those accessible for review, are primarily in Western languages, influencing the representation of projects in this analysis and setting the scope of the article.

All of the presented citizen science quantum games have been discontinued and do not have any recent updates on their progress, though some of them are still playable. Some of these projects have run out of resources related to time, funding, manpower, or  because the research aim of the group has shifted \cite{wootton2017,scienceathomedocumentary,piispanen2023projects, piispanen2024}. These citizen science games on quantum physics were able to reach millions of players world-wide. The public interest towards quantum technologies, particularly quantum computers and quantum computing, has been observed to have increased since the most active years of these games \cite{fox2020, bitzenbauer2021}. This might point towards a fruitful ground for future quantum citizen science games.\\

Within this context, games have successfully been employed in the pursuit of science research in areas like astronomy, taxonomy, biomedical imaging and protein folding, but so far citizen science games in quantum physics have not stayed viable, though the developers express their trust in the future of citizen science quantum games \cite{devitt2016,scienceathomegames,piispanen2024}. The aforementioned citizen science games in biological sciences and astronomy have primarily relied on human pattern recognition, spatial reasoning and the ability of participants to name and characterise data based on images \cite{cooper2011phd,vohland2021}. However, when delving into the realm of quantum physics, a stark contrast emerges. We acknowledge that the average citizen possesses minimal knowledge, if any, about quantum physics, which may challenge reaching and motivating suitable players for a citizen science game. Furthermore, there is a notable absence of visual references to quantum mechanical phenomena in our daily experiences. In this domain, our understanding is primarily derived from mathematical formalism, simulations and visual representations of the quantum world based on these. One could even argue that scientists themselves sometimes struggle to conceptualise the visual aspects of `what quantum looks like' when communicating to the general public. 

Undoubtedly, this presents the most formidable challenge in developing citizen science games within the realm of quantum sciences — the delicate balance between a scientifically accurate representation of phenomena and the creation of a game interface that is both intuitive and approachable. The researchers behind the games \textit{meQuanics}, \textit{Quantum Moves}, and \textit{QWiz} were originally inspired to create games that would allow the gaining of intuition about quantum physical phenomena, which would then lead to human-sourced solutions distinct from the solutions reached with algorithmic methods \cite{mequanicssab, mequanicsjacob}. This hope for intuition growth reflects how John Preskill envisioned future generations in a quote sometimes used to motivate the design of quantum games \begin{quote}
\textit{“Perhaps kids who grow up playing quantum games will acquire a visceral understanding of quantum phenomena that our generation lacks. Furthermore, quantum games could open a niche for quantum machine learning methods, which might seize the opportunity to improve gameplay in situations where quantum entanglement has an essential role}" \cite{preskill2018}.
\end{quote} Today, we see that the development of citizen science games tends to lean towards the interplay of human and artificial intelligence, such as machine learning algorithms \cite{franzen2021,rafner2021,rafner2022,sherson2023}.

Furthermore on the dimensions of quantum games, it is common for quantum games to have a perceivable dimension of quantum physics \cite{piispanen2022}, but particularly for citizen science quantum games, it is of course not necessary for the player to be aware of the technicalities related to form a connection between the visuals and the problem itself. Still, emphasis has been placed on establishing a clear connection to quantum physics research in all the material related to the presented citizen science quantum games, as this has been thought to play a pivotal role in motivating players to engage with and revisit these games \cite{lieberoth2014}. For educational games, on the other hand, the perceivable dimension is pivotal. Perceivable elements constitutes not only of visual queues, but also of aspects that are perceivable through the gameplay such as game rules, mechanics, and storyline \cite{piispanen2022}.\\

Upon the definition of quantum games, serious uses of quantum games were aligned along the \textit{dimension of scientific purposes} (see Table \ref{table:definition}) \cite{piispanen2022}. It was proposed that the dimension of scientific purposes of quantum games could be further dissected to separate citizen science objectives from educational objectives in order to bring rationalised structure both to the development of citizen science games as well as educational games \cite{piispanen2022}. Certain differences between the objectives of these games support this proposal. For the presented quantum citizen science games, the problem at hand is of a very specific kind, whereas for the development of educational games the purpose often is to provide the player with a general understanding of more basic quantum mechanical phenomena. It is also more straightforward to confirm whether an educational game teaches a topic than it is to develop a citizen science game solving a problem with unknown solution spaces.

Revision is an important aspect of educational games, whether it is done within the game or with in-classroom supervision. For citizen science games the quality of the produced data is more important to the scientist than the player. The player may even be provided with an artificial reward-system. Moreover, while for the player of educational games a single challenge, level or a puzzle might be crucial for the learning of the overall concept addressed in the game, for a citizen science game the first few levels are often implemented only as tutorials for understanding the game mechanics and only the data from the higher levels serve the related research objectives, like in the game \textit{QWiz} \cite{piispanen2024}. The aim of the tutorials is to provide intuition about the underlying problem but more so for the interface of the game.  

\section{Future Directions of Citizen Science Quantum Games}
For us to understand \textit{how} citizen science quantum games are designed, it is not enough simply to study the games as artefacts. Collaboration and open discussion on best practices in the development process are equally important. So far, the academic literature surrounding the designing and development process of citizen science quantum games is almost non-existent with only a couple of exceptions that offer design guidelines for developing quantum games with a particular focus on those connecting to numerical simulations \cite{piispanen2023projects,piispanen2024}. These guidelines emphasise an early involvement of both game design and quantum physics expertise, early design decisions on including a citizen science prospective and underline the importance of providing for visual communication. In addition, the learnings from the series of Quantum Game Jams serve as important indicators not only for organising science game jams, but provide strong guide for any serious game development project \cite{piispanen2023qgj}.

Experiences and lessons on quantum game development have also been shared from the perspective of a professional game designer and artist \cite{archer2022} and by interdisciplinary teams developing educational quantum games \cite{ashoori2018, entanglion,anupam2020}. More generally, various actors, both academic and industrial, have offered directions for quantum game development from the perspective of building game rules on quantum physical principles and on the practicalities related to using \textit{Qiskit} in a game \cite{wootton2017howto, becker2019, cantwell2019,wootton2021}. In addition to these, the dimensions of quantum games may prove helpful when designing a meaningful connection of a game to quantum physics especially for serious game design \cite{piispanen2022}.\\

From the interactive citizen science platform of galaxy classification based on pictures, \textit{Galaxy Zoo} has evolved into the multidisciplinary citizen science project platform \textit{Zooniverse}, which allows the creation of citizen science projects based on existing data \cite{raddick2009citizen,simpson2014}. A similar joint effort of resources could be an interesting initiative for the field of citizen science quantum games. Alternatively, focusing specifically on quantum optimal control related research topics could create a fruitful foundation for a collaborative citizen science platform on optimal control problems in general. Optimal control problems, oversimplifying somewhat, are addressed by identifying a suitable criterion for what is considered optimal and creating a measure for it \cite{kirk2004optimal}. Therefore optimisation problems offer a possibility to communicate something about the underlying system through a metric, which is transferable to a visual form understandable by a human-being. Any metric might make the creation of a visual interface more straightforward, but poses the question of the most suitable metric for measuring the suitability of an answer. An interesting topic of research would be to study what are the characteristics of complex (numerical) research problems, as in quantum physics research, that indicate the usefulness of visual representations and thus allow for human-sourced solutions through a citizen science platform like a game.

There is some indication that humans might possess the skill to learn intuition on abstract complex optimisation systems and on solving quantum optimisation problems by interacting with a visual representation that is based on a numerical simulation \cite{malaspina2010,heck2018}. It would be intriguing to study what humans are able to discern when interacting directly with quantum computer-sourced data in games for the purpose of building intuition. Maybe in the future we might have a platform for quantum physics related citizen science projects in a way that would allow the use of quantum computers or the hybrid use of quantum computers and supercomputers. So far the obstacle has been the limited access to quantum computers. Possibly a tournament-type event with first rounds on tutorial games or tasks with classically sourced numerical simulations could both educate players on the logic of the game or task, but also eliminate players that would not have the motivation for longer term involvement in the citizen science research mission. This could provide a system where a limited number of players get to test out their abilities on the levels of the designed game that require access to a quantum computer and could additionally allow allocating the needed computational resources to be available during the time of final rounds. \textit{The BIG Bell Quest} was carried out in a manner like this regarding the available laboratories. 

We may of course question the relevance of a connection to a quantum computer to build intuition about the behaviour of the studied system, as numerical simulations have shown to suffice in the case of \textit{Quantum Moves 2}. But a single game and a single scenario cannot give the full picture. By performing a comparative study on the use of numerically simulated data on quantum systems against source data from a quantum computer on different types of tasks could bring us more insight on whether a person without prior knowledge about quantum physics could develop intuition about quantum physics problems and if the use of quantum computers bring any extra value. More research into the idea could also lead us towards visualisations and user interfaces for quantum software that require no specific technical understanding of quantum phenomena from the end user.

For the development of quantum games with serious purposes, the user interface needs to provide scientifically accurate representations. Though it is understandably highly important to ensure that the visual representation of quantum physical phenomena in educational games is scientifically accurate and provides for a well-founded aid towards the learning objectives, for citizen science games there might be more freedom. The art of finding the middle-ground of a scientifically accurate representation and intuitively motivating interface is an interesting and challenging objective for future research, specifically on citizen science quantum games. Once we have a solid understanding of visual communication of quantum phenomena and/or have a more quantum literate general public, the development process of citizen science quantum games will have more concrete design decisions related to what will be perceived through the game.\\

Whatever the future of citizen science quantum games or citizen science projects related to quantum physics, it should definitely take advantage of collaboration and the open sharing of resources. Equally important to learning about success stories like \textit{Foldit} and \textit{Galaxy Zoo} is learning about possible obstacles in projects that ended up being discontinued \cite{wootton2017,piispanen2023projects,piispanen2024}. A successful project combining game design, top quality research and aspects of citizen science needs a long-term plan, adequate allotting and allocation of resources as well as expertise from all the respective fields.

\section{Conclusions}
In this article, we have traced the evolution of quantum physics-related games, commonly known as \textit{quantum games} according to the definition in \cite{piispanen2022}, and explored their intersection with citizen science. We see that the landscape of quantum games has evolved under three driving forces: serious game use, the development of quantum computers and the influence of open game developing events such as \textit{Quantum Game Jams}. The article presents eleven citizen science quantum games and game prototypes, one gamified citizen science platform for crowdsourcing, and an interactive art installation used for citizen science. All of these projects are discontinued and only a couple of them provide insight on the game development process. More citizen science game prototypes have been developed at Quantum Game Jam events.

Nevertheless, the interest in understanding and utilising quantum technologies persists, suggesting a fruitful ground for the use of quantum games. However, much remains to be uncovered, especially regarding how the visual representation of quantum physics best correlates with the user interfaces of these games. Transparently sharing resources, project reports, and insights is strongly recommended for the ongoing advancement of the field of citizen science quantum games. For the future of quantum games and citizen science quantum games, this paper points to external references with development guidelines.

\backmatter



\bmhead{Data Availability Statement}
All data that support the findings of this study are either included within the article or derived from the open-source listing of \textit{The List of Quantum Games} at \cite{quantumgames}.

\bmhead{Acknowledgments}
The author would like to express her gratitude to MSc. Daria Anttila for our constructive discussions regarding the structure and content of the article. The author gratefully acknowledges that her research has been funded by the Väisälä Foundation and the Alfred Kordelin Foundation.\\



\bibliography{sn-bibliography}

\end{document}